\documentclass[aps,prd,preprint,groupedaddress]{revtex4}
\usepackage{graphicx}% Include figure files

\begin{document}

%\preprint{}

\title{Light-flavor sea-quark distributions in the nucleon\\
in the SU(3) chiral quark soliton model (I) \\
--- phenomenological predictions ---}


\author{M.~Wakamatsu}
\email[]{wakamatu@miho.rcnp.osaka-u.ac.jp}
%\thanks{}
\affiliation{Department of Physics, Faculty of Science, \\
Osaka University, \\
Toyonaka, Osaka 560, JAPAN}


%\date{\today}

\begin{abstract}
Theoretical predictions are given for the light-flavor sea-quark
distributions in the nucleon including the strange quark ones on the
basis of the  flavor SU(3) version of the chiral quark soliton model.
Careful account is taken of the SU(3) symmetry breaking effects 
due to the mass difference $\Delta m_s$ between the strange and 
nonstrange quarks, which is the only one parameter necessary for the 
flavor SU(3) generalization of the model.
A particular emphasis of study is put on the {\it light-flavor
sea-quark asymmetry} as exemplified by the observables
$\bar{d} (x) - \bar{u} (x), 
\bar{d} (x) / \bar{u} (x), \Delta \bar{u} (x) - \Delta \bar{d} (x)$
as well as on the {\it particle-antiparticle asymmetry}
of the strange quark distributions represented by
$s (x) - \bar{s} (x), s (x) / \bar{s} (x),
\Delta s (x) - \Delta \bar{s} (x)$ etc.
As for the unpolarized sea-quark distributions, the predictions of the
model seem qualitatively consistent with the available phenomenological
information provided by the NMC data for $\bar{d} (x) - \bar{u} (x)$,
the E866 data for $\bar{d} (x) / \bar{u} (x)$, the CCFR data and 
Barone et al.'s fit for $s (x) / \bar{s} (x)$ etc. The model is 
shown to give several unique predictions also for the spin-dependent 
sea-quark distribution, such that $\Delta s (x) \ll \Delta \bar{s}(x)
\lesssim 0$ and $\Delta \bar{d}(x) < 0 < \Delta \bar{u}(x)$,
although the verification of these predictions must await more
elaborate experimental investigations in the near future.
\end{abstract}

%\keywords{}

\maketitle

%\section{}
%\subsection{}
%\subsubsection{}

\section{Introduction}

As is widely known, the perturbative QCD can predict only the 
$Q^2$-dependence of parton distribution functions (PDF), whereas
it can say nothing about the PDF at a prescribed energy scale. 
To predict PDF themselves, we need to solve nonperturbative
QCD, which is an extremely difficult theoretical problem.
It cannot be denied that, at least at the present stage,
we cannot be too much ambitious in this respect.
Still, we can do qualitatively interesting investigations.
The key observation here is the following.
In their semi-phenomenological analyses of PDF, Gl\"{u}ck,
Reya and Vogt prepared the initial PDF at fairly low energy scale
around 600 MeV, in contrast to the standard consent of perturbative
QCD, and they concluded that light-flavor sea-quark (or antiquark)
components are absolutely necessary even at this relatively
low energy scale \cite{GRV95},\cite{GRSV96}.
Furthermore, even the flavor asymmetry of the sea-quark distributions
have been established by the celebrated NMC measurement \cite{NMC91}.
The origin of this sea-quark asymmetry seems definitely
nonperturbative, and cannot be explained by the sea-quarks
radiatively generated through the perturbative QCD evolution processes.
Here we certainly need some low energy (nonperturbative) mechanism
which generates sea-quark distributions in the nucleon.
In our opinion, the chiral quark soliton model (CQSM) is the 
simplest and most powerful effective model of QCD, which fulfills 
the above physical requirement \cite{DPP88}\nocite{WY91}\nocite{W92P}\nocite{CBKPWMAG96}\nocite{ARW96}--\cite{DP01}.
Although it may still be a toy model in the sense that the gluon
degrees of freedom are only implicitly
handled, it has several nice features that are not shared by 
other effective models like the MIT bag model.
Among others, most important in the above-explained context is 
its field theoretical nature, i.e. the proper account of the
polarization of Dirac sea quarks, which enables us to make reasonable
estimation not only of quark distributions but also of
{\it antiquark distributions}
\cite{DPPPW96}\nocite{DPPPW97}\nocite{WK98}--\cite{WK99}.
It has already been shown that,
{\it without introducing any adjustable 
parameter}, except for the initial-energy scale of the $Q^2$-evolution,
the CQSM can describe nearly all the qualitatively noticeable features 
of the recent high-energy deep-inelastic scattering observables.
It naturally explains the NMC observation, i.e. the excess of
$\bar{d}$-sea over the $\bar{u}$-sea
in the proton \cite{WK98},\cite{W91A}\nocite{W91B}--\cite{W92},
\cite{PPGWW99}. It also reproduces the characteristic features
of the observed longitudinally polarized structure functions
of the proton, the neutron and the deuteron \cite{WK99},\cite{WW00A}.
Even the most puzzling 
observation, i.e. the unexpectedly small quark spin fraction
of the nucleon, can be explained at least qualitatively with 
no need of a large gluon polarization at the low renormalization
scale \cite{WY91},\cite{WW00B}.
Finally, the model predicts a sizably large isospin asymmetry 
also for the spin-dependent sea-quark distributions, which we
expect will be confirmed by near future experiments
\cite{DPPPW96},\cite{WK99},\cite{WW00A},\cite{DGPW00}.

The above-mentioned unique feature of the CQSM is believed to 
play important roles also in the study of hidden strange 
quark excitations in the nucleon, which entirely have {\it non-valence
character} \cite{W02}.
The main purpose of the present study is to give 
theoretical predictions for both of the unpolarized and the 
longitudinally polarized strange quark distributions in the nucleon, 
on the basis of the CQSM generalized to the case of flavor SU(3).
Naturally, because of fairly large mass difference between the 
strange and nonstrange quarks, the flavor SU(3) symmetry is not 
so perfect symmetry as the flavor SU(2) one is.  We must take 
account of this symmetry breaking effects in some way 
or other.
Here, we shall accomplish it relying upon the first order perturbation 
theory. It should be emphasized that, in our effective theory at quark
level, this effective mass difference $\Delta m_s$ between the $s$-quark
and the $u,d$-quarks is the {\it only one additional parameter} necessary
for the flavor SU(3) generalization of the CQSM.
Through the study outlined above, we will be able to answer
several interesting questions as follows.
How important in nature is the admixture or the virtual 
excitation of $s$-$\bar{s}$ pairs in the nucleon,
a system of total strange-quantum-number being zero?
Does the asymmetry of the $s$-quark and $\bar{s}$-quark distributions 
exist at all? If it exists, how large is it? Do we expect an 
appreciable particle-antiparticle asymmetry also for the 
spin-dependent strange quark distribution? 
We also want to verify whether a favorable prediction of the 
flavor SU(2) CQSM, i.e. the excess of the $\bar{d}$-sea over 
the $\bar{u}$-sea in the proton, is taken over by the SU(3) model
or not. What answer do we obtain for the isospin asymmetry of the 
spin dependent sea-quark distributions $\Delta \bar{u} (x) 
- \Delta \bar{d} (x)$ in the flavor SU(3) CQSM?

In consideration of the length of the theoretical formulation of the
model, we think it more appropriate to organize the paper as follows.
That is, for the benefit of readers who have interest only in the
phenomenological consequences of the SU(3) CQSM, we leave the description
of the full theoretical formalism to a separate paper. (This paper will
hereafter be referred to as II.) Instead, we will
give in next section a brief summary of what dynamical assumptions
the model is constructed on and what approximations are necessary there.
Next, in sect.3, we compare the theoretical
predictions of the model with available phenomenological information.
We shall also give some discussions on the physical origin of the unique 
predictions of the model as to the light-flavor sea-quark asymmetry. 
Finally, in sect.4, some concluding remarks will made.

\vspace{4mm}
\section{A brief summary of the model}

Since the flavor $SU(3)$ CQSM is constructed on the basis of the
flavor $SU(2)$ model, we first recall some basics of the $SU(2)$ CQSM.
It is specified by the effective Lagrangian \cite{DPP88},
\begin{equation}
 {\cal L} \ = \ \bar{\psi} (x) (\,i \! \not\!\partial - M U^{\gamma_5} (x)
 \,) \psi (x) .
\end{equation}
with 
\begin{equation}
 U^{\gamma_5} (x) \ = \ e^{i \gamma_5 \pi (x) / f_{\pi}}, \hspace{15mm}
 \pi (x) = \pi_a (x) \tau_a \ \ \ (a = 1, \cdots, 3)
\end{equation}
which describes the effective quark fields with a dynamically
generated mass $M$, interacting with massless
pions. The nucleon (or $\Delta$) in this model appears as a
rotational state of a symmetry-breaking hedgehog object, which
itself is obtained as a solution of the self-consistent Hartree problem
with infinitely many Dirac-sea quarks \cite{DPP88},\cite{WY91}.
The theory is not renormalizable, and it is defined with
an ultraviolet cutoff. In the Pauli-Villars regularization
scheme, which is used throughout the present analysis,
that which plays the role of the ultraviolet cutoff is the Pauli-Villars
mass $M_{PV}$ obeying the relation
$\left( N_c M^2 / 4 \pi^2 \right) \ln {\left( M_{PV} / M \right)}^2
=  f_\pi^2$ with $f_\pi$ the pion weak decay constant \cite{DPPPW96}.
Using the value $M \simeq 375 \,\mbox{MeV}$, which is obtained
from the phenomenology of nucleon low energy observables,
this relation fixes the Pauli-Villars mass as
$M_{PV} \simeq 562 \,\mbox{MeV}$.
Since we are to use these values of $M$ and $M_{PV}$, there is
{\it no free parameter} additionally introduced into the calculation of
distribution functions \cite{WW00A}.

The basic lagrangian of the SU(3) CQSM is given as 
\begin{equation}
 {\cal L} \ = \ \bar{\psi} (x) (\,i \! \not\!\partial - M U^{\gamma_5} (x)
 - \Delta m _s P_s \,) \psi (x) .
\end{equation}
with 
\begin{equation}
 U^{\gamma_5} (x) \ = \ e^{i \gamma_5 \pi (x) / f_{\pi}}, \hspace{15mm}
 \pi (x) = \pi_a (x) \lambda_a \ \ \ (a = 1, \cdots, 8)
\end{equation}
and 
\begin{equation}
 \Delta m_s P_s \ = \ \Delta m_s \,
 \left( \frac{1}{3} - \frac{1}{\sqrt{3}} \,\lambda_8 \right) \ = \ 
 \left(
 \begin{array}{ccc}
 0 & 0 & 0 \\
 0 & 0 & 0 \\  
 0 & 0 & \Delta m_s
 \end{array}
 \right)
\end{equation}
It is a straightforward generalization of the SU(2) model \cite{WAR92},
\cite{BDGPPP93}, except for one important new feature, i.e. the
existence of SU(3) symmetry breaking term due to the sizably large mass
difference $\Delta m_s$ between the strange and nonstrange quarks.
This mass difference $\Delta m_s$ is the only one 
additional parameter necessary for the flavor SU(3) generalization of 
the CQSM.

Now, the fundamental dynamical assumption of the SU(3) CQSM is as follows.
The first is the embedding of the SU(2) self-consistent mean-field (of 
hedgehog shape) into the SU(3) matrix as
\begin{equation}
 U^{\gamma_5}_0 (\mbox{\boldmath $x$}) \ = \ 
 \left( \begin{array}{cc}
 {e^{i \gamma_5 \mbox{\boldmath $\tau$} \cdot 
 \hat{\mbox{\boldmath $r$}} F(r)}} & 0 \\
 0 & 1
 \end{array}
 \right) \, ,
\end{equation}
just analogous to the SU(3) Skyrme model \cite{G84},\cite{MNP84}.
The next assumption is the semiclassical quantization of the rotational 
motion in the SU(3) collective coordinate space represented as 
\begin{equation}
 U^{\gamma_5} (\mbox{\boldmath $x$}, t) \ = \ 
 A (t) \, U^{\gamma_5}_0 (\mbox{\boldmath $x$}) \,A^\dagger (t) ,
\end{equation}
with
\begin{equation}
 A(t) \ = \ e^{-i \Omega t}, \hspace{15mm}
 \Omega \ = \ \frac{1}{2} \Omega_a \lambda_a \ \in \ SU(3) .
\end{equation}
The semiclassical quantization of this collective rotation leads to a 
systematic method of calculation of any nucleon observables, including 
the parton distribution functions, which is given as a perturbative series 
in the collective angular velocity operator $\Omega$.
(We recall that this reduces to a kind 
of $1 / N_c$ expansion, since $\Omega$ itself is an order $1/N_c$ quantity.) 
In the present study, all the terms up to the first order in $\Omega$ are
consistently taken into account, basically according to the path integral 
formalism explained in \cite{WK99}. Unfortunately, in the evaluation of the 
$O(\Omega^1)$ contribution to the parton distribution function, we
sometimes encounter ordering ambiguity of the collective space operators
\cite{WW93}\nocite{CBGPPWW94}\nocite{SW95A}\nocite{SW95B}\nocite{W95}\nocite{W96}\nocite{CGP95}\nocite{AW93}--\cite{PWG99}.
In the case of flavor SU(2) CQSM, this ordering ambiguity of two
collective operators is known to be avoided by adopting the
time-order-preserving collective quantization procedure, which leads
to the resolution of the long-standing $g_A$ problem inherent in the
soliton model based on the classical hedgehog
configuration \cite{WW93},\cite{CBGPPWW94}.
However, it was pointed out that there exists some inconsistency between
this particular quantization procedure and the basic dynamical framework of
the SU(3) CQSM, i.e. the embedding of the SU(2) hedgehog mean-field into
the SU(3) matrix followed by the subsequent quantization of the rotational
motion in the full SU(3) collective coordinate space \cite{PWG99}.
Here, we avoid this problem simply following the symmetry preserving
approach advocated in \cite{PWG99}. The more detailed discussion of this
delicate problem will be given in II.

  Another important feature of the model lagrangian is the existence of 
sizably large SU(3) breaking term. We assume that the SU(3) symmetry breaking 
effects can be estimated by using the first order perturbation theory in the 
parameter $\Delta m_s$. In fact, its perturbative treatment would be justified 
(though not completely), since the effective mass difference $\Delta m_s$ 
of the order of 100 MeV is much smaller than the typical energy scale of 
the model, which may be specified by the Pauli-Villars cutoff mass around 
600 MeV. In the present investigation, we are to take account of three
possible SU(3) breaking corrections in a consistent way, which are all first
order in the mass parameter $\Delta m_s$. The simultaneous account of these
corrections is shown to be essential for maintaining the quark number sum
rules for the unpolarized distribution functions.
The detail will again be explained in II.

\vspace{4mm}
%\newpage
\section{Numerical Results and Discussions}

The basic lagrangian of the model contains three physical
parameters, the weak pion decay constant $f_\pi$, the
dynamically generated effective quark mass $M$, and the
mass difference $\Delta m_s$ between the strange and
nonstrange quarks. As usual, $f_\pi$ is fixed to be its
physical value, i.e. $f_\pi = 93 \,\mbox{MeV}$.
On the other hand, $M$ is taken to be $375 \,\mbox{MeV}$,
which is the same value as used in our previous analysis of
the nucleon spin structure functions within the framework
of the flavor SU(2) CQSM \cite{WK99}.
As a consequence, only one parameter
remains in the SU(3) CQSM : it is $\Delta m_s$, i.e. the
effective mass difference between the strange and nonstrange
quarks.
In the present analysis, we have tried to vary this parameter
within the physically reasonable range, i.e.
$60 \,\mbox{MeV} < \Delta m_s < 170 \,\mbox{MeV}$, and
found that overall success of the theory is obtained for the
value of $\Delta m_s$ around $100 \,\mbox{MeV}$. 
All the following analyses are thus carried out by using the
value $\Delta m_s = 100 \,\mbox{MeV}$.

The model contains ultraviolet divergences so that it must be
regularized by introducing some physical cutoff. Following the
previous studies, we use the Pauli-Villars regularization scheme.
In this scheme, any nucleon observables including quark distribution
functions in the nucleon are regularized through the subtraction :
\begin{equation}
 {\langle O \rangle}^{reg} \ \equiv \ {\langle O \rangle}^M \ - \ 
 {\left( \frac{M}{M_{PV}} \right)}^2 \,{\langle O \rangle}^{M_{PV}} .
 \label{PVreg}
\end{equation}
Here ${\langle O \rangle}^{M}$ denotes the nucleon matrix element of
an operator $O$ evaluated with the original effective action with the
mass parameter $M$, while ${\langle O \rangle}^{M_{PV}}$ stands for the
corresponding matrix element obtained from ${\langle O \rangle}^{M}$
by replacing the parameter $M$ with the Pauli-Villars cutoff mass
$M_{PV}$. We emphasize that the Pauli-Villars mass $M_{PV}$ is not
an adjustable parameter of the model.
Demanding that the regularized action reproduces
the correct normalization of pion kinetic term in the corresponding
bosonized action, $M_{PV}$ is uniquely fixed by the relation
\begin{equation}
 \frac{N_c}{4 \,\pi^2} \,M^2 \,\log \,\frac{M_{PV}^2}{M^2} \ = \ 
 f_\pi .
\end{equation}
For $M = 375 \,\mbox{MeV}$, this gives $M_{PV} \simeq 562 \,
\mbox{MeV}$.

Several additional comments are in order for the regularization scheme
explained above. First, in the present investigation, the
regularization specified by (\ref{PVreg}) is introduced into all
the observables, including those related to the imaginary part of the
Euclidean action. This is in contrast to some authors' claim that
the imaginary action should not be regularized \cite{BDGPPP93},
\cite{WRG99}. The ground of their assertion is that the imaginary
part of the Euclidean action is ultraviolet finite and that the
introduction of regularization would destroy conservation laws of
some fundamental quantities like the baryon number and/or the
quark numbers.
It would be true if one uses the energy cutoff scheme like the
proper-time regularization scheme. If fact, the proper-time
regularization scheme is known to lead a violation of
baryon-number conservation law at the level of $3 \,\%$.
This is not the case for the Pauli-Villars regularization scheme,
however.
The baryon-number is just intact in this regularization scheme.
Generally speaking, the
introduction of regularization would give some effects on the quark
distribution functions even though the fundamental conservation laws
are intact. Since what we are handling is not a renormalizable theory
but an effective theory, a different choice of regularization scheme
leads to a different effective theory. We can say that our effective
theory is defined with the above-explained regularization
prescription.

Secondly, as was shown in Ref.~\cite{KWW99},
the Pauli-Villars scheme with
a single subtraction term is not a completely satisfactory
regularization procedure. It fails to remove ultraviolet divergences
of some special quantities like the vacuum quark condensate, which
contains quadratic divergence instead of logarithmic one.
For obtaining finite answers
also for these special observables, the single-subtraction
Pauli-Villars scheme is not enough.
It was shown that more sophisticated Pauli-Villars
scheme with two subtraction terms meets this requirement \cite{KWW99}.
Fortunately, the self-consistent solution of the CQSM obtained in this
double-subtraction Pauli-Villars scheme is only slightly different
from that of the naive single-subtraction scheme,
except when dealing with some special quantities containing quadratic
divergences \cite{KWW99}.
Considering the fact that the calculation of
quark distribution functions in the CQSM is extremely time-consuming
and that the most nucleon observables are rather insensitive to
which regularization scheme is chosen, we shall simply use here
the single-subtraction Pauli-Villars scheme
except for one special quantity to be just mentioned.
It is the quantity
$\bar{\sigma}$ defined in Eq.(206) of II, i.e. the scalar charge of
the nucleon. This parameter, appearing in the representation mixing
$\Delta m_s$ correction to the quark distribution functions, contains a
quadratic divergence that can be regularized only by using the
double-subtraction Pauli-Villars scheme. Since the predictions of the
double-subtraction Pauli-Villars scheme for this quantity is not so
far from the canonical value $\bar{\sigma} \simeq 7.5$,
which is obtained from the analysis of the pion-nucleon sigma
term \cite{GLS91}, we shall simply use this value in the present
study.

To compare the predictions of the CQSM with the existing high energy
data, we must take account of the scale dependencies of the quark
distribution functions. This is done by using the Fortran codes
provided by Saga group \cite{MK96},\cite{HKM98A},\cite{HKM98B}
which enable us to solve the DGLAP equations
at the next-to-leading order. The initial
energy of this scale evolution is fixed to be the value
$Q^2 = 0.30 \,\mbox{GeV}^2$ throughout the whole investigation.
Strictly speaking, it is a serious question how much meaning one can
give to starting the QCD evolution at such low energy as
$Q^2 \simeq (600 \,\mbox{MeV})^2$, even though it is just motivated
by a similar semi-phenomenological prescription by Gl\"{u}ck, Reya
and Vogt \cite{GRV95}. Furthermore, any precise statement about the
model energy scale, where the $Q^2$-evolution of the theoretical
PDF should be started, is hard to give at our present understanding
of the effective theory within the full QCD framework.
(Worthy of special mention here is an interesting challenge to this
difficult problem \cite{BJY02}.) One should then keep in mind
theoretical uncertainties introduced by such a semi-phenomenological
prescription.

Generally, the theoretical distribution functions obtained in the
CQSM have unphysical tails beyond $x = 1$, although they are very
small in magnitude.
These unphysical tails of the theoretical distributions come from an
approximate nature of our treatment of the soliton center-of-mass
motion, which is essentially nonrelativistic.
Since the Fortran programs of Saga
group require that the distribution functions must vanish exactly for
$x \geq 1$, we introduce a $x$-dependent cutoff factor $(1 - x^{10})$
into all the theoretical distribution functions prepared at
the model energy scale before substituted into the DGLAP equations.
One can however confirm
from Fig.~13 of Ref.~\cite{WK99} that the introduction of this cutoff
factor hardly modifies the original distributions except for
their tail behavior near and beyond $x = 1$.

\renewcommand{\textfraction}{0.1}
\begin{figure}[htb] \centering
\begin{center}
 \includegraphics[width=15.0cm]{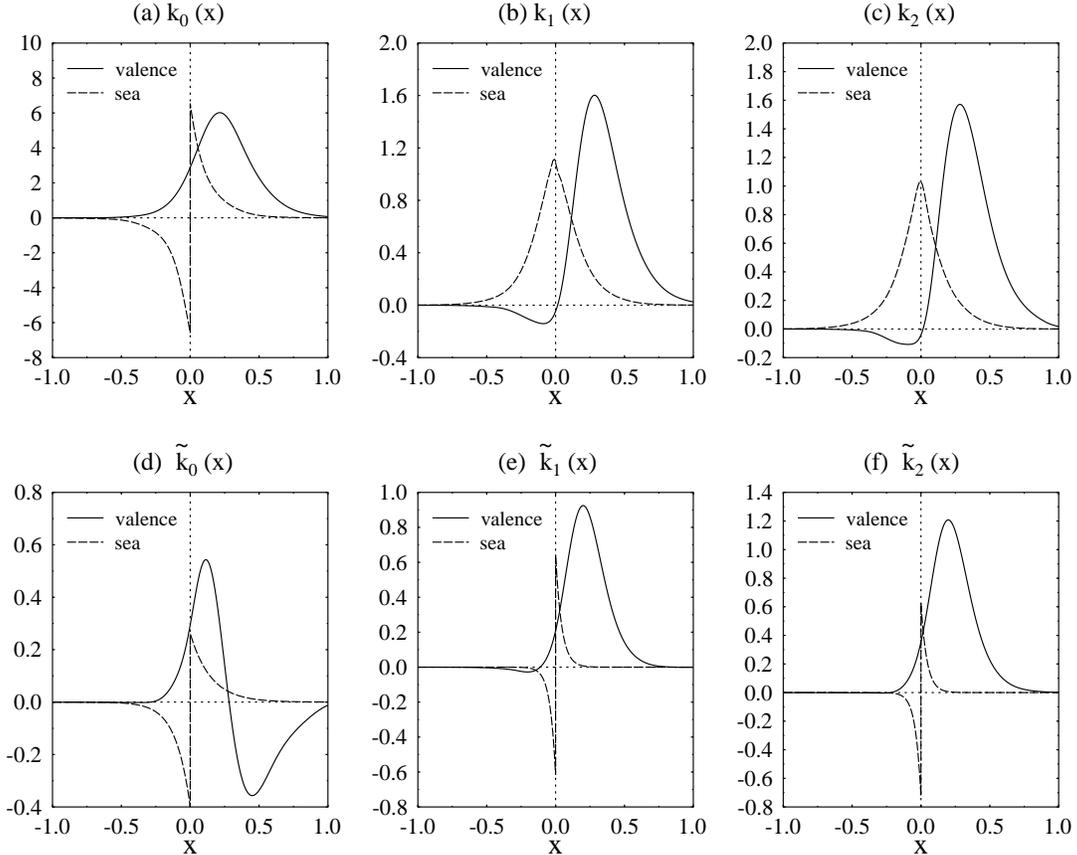}
\end{center}
\vspace*{-0.5cm}
\renewcommand{\baselinestretch}{1.20}
\caption{Six basic functions necessary for obtaining
unpolarized distribution functions within the SU(3) CQSM
with $\Delta m_s$ corrections. Here, the solid and dashed curves
respectively stand for the contributions of $N_c$ valence quarks
and those of Dirac-sea quarks. \label{unpolmodel1}}
\end{figure}

Now we are ready to show the results of the numerical calculations.
Fig.~\ref{unpolmodel1} shows six basic functions necessary for
evaluating unpolarized distribution functions.
Here, the first three functions
$k_0 (x), k_1 (x)$ and $k_2 (x)$ appear in the leading
$O(\Omega^0 + \Omega^1)$ contributions, while the remaining three
functions $\tilde{k}_0 (x), \tilde{k}_2 (x)$ and $\tilde{k}_2 (x)$
are contained in the SU(3) breaking corrections to the unpolarized
distribution functions.
In all the figures, the solid and dashed curves represent
the contributions
of the $N_c$ valence quarks and those of the Dirac-sea quarks.
(We recall that the terminology ``valence quark'' above should not be
confused with the corresponding term in the quark-parton model.
The valence quarks in the CQSM denote quarks occupying the particular
bound-state orbital, which emerges from the positive-energy Dirac
continuum under the influence of the background pion field of
hedgehog shape. Note however that the
valence quark distribution in the sense of quark-parton model is
easily obtained, as a difference of quark and antiquark
distributions evaluated in the CQSM.
One clearly sees that the effects of Dirac-sea quarks, or equivalently
the vacuum polarization effects, are very important in all the basic
distribution functions shown in Fig.~\ref{unpolmodel1}.
One can also convince that the above-mentioned unphysical tails of
the distribution functions beyond $x = 1$ are really very small
and of little practical importance.

\begin{figure}[htb] \centering
\begin{center}
 \includegraphics[width=15.0cm]{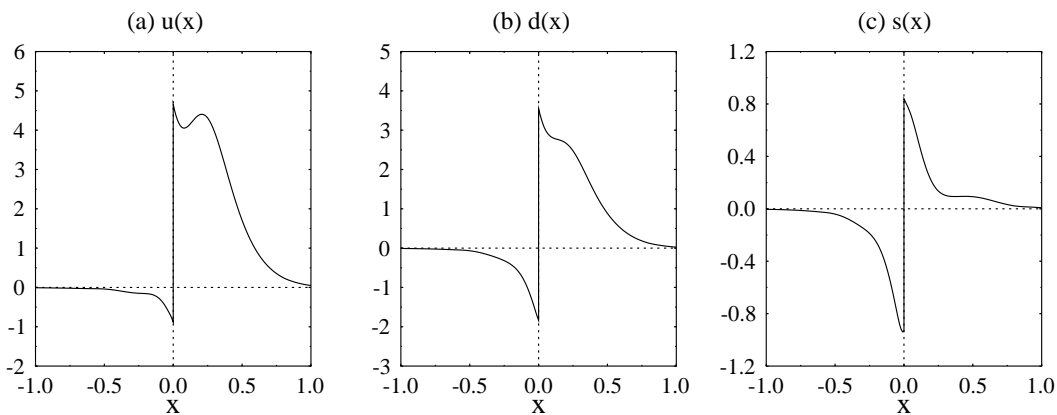}
\end{center}
\vspace*{-0.5cm}
\renewcommand{\baselinestretch}{1.20}
\caption{The unpolarized distribution functions with respective
flavors. The distribution functions in the negative $x$ region should
actually be interpreted as the antiquark distributions
according to the rule $q(-x) = - \,\bar{q}(x)$ with
$0 < x < 1$. \label{unpolmodel2}}
\end{figure}

By using these basic functions, we can calculate any unpolarized
distribution function with a specified flavor.
Shown in Fig.~\ref{unpolmodel2}
are the theoretical unpolarized distribution functions corresponding
to three light flavors $u, d$ and $s$.
Remember that a distribution functions in the
negative $x$ region are interpreted as antiquark distributions
according to the rule $q(-x) = - \,\bar{q}(x)$
with $0 < x < 1$.
The familiar positivity constraint for the unpolarized quark and
antiquark distributions means that $q(x) > 0$ for $0 < x < 1$,
while $q(x) < 0$ for $-1 < x < 0$.
One clearly sees that our theoretical calculation
legitimately satisfies this general constraint for the PDF.
One can understand that this is not a trivial result, if one remembers
the fact that the previous calculations by T\"{u}bingen group carried
out in the so-called ''valence-quark-only'' approximation violate this
general constraint in an intolerable way \cite{SRW99}.
This proves our assertion that the
proper account of the vacuum polarization contributions is vital to
give any reliable predictions for the sea-quark distributions.

\begin{figure}[htb] \centering
\begin{center}
 \includegraphics[width=15.0cm]{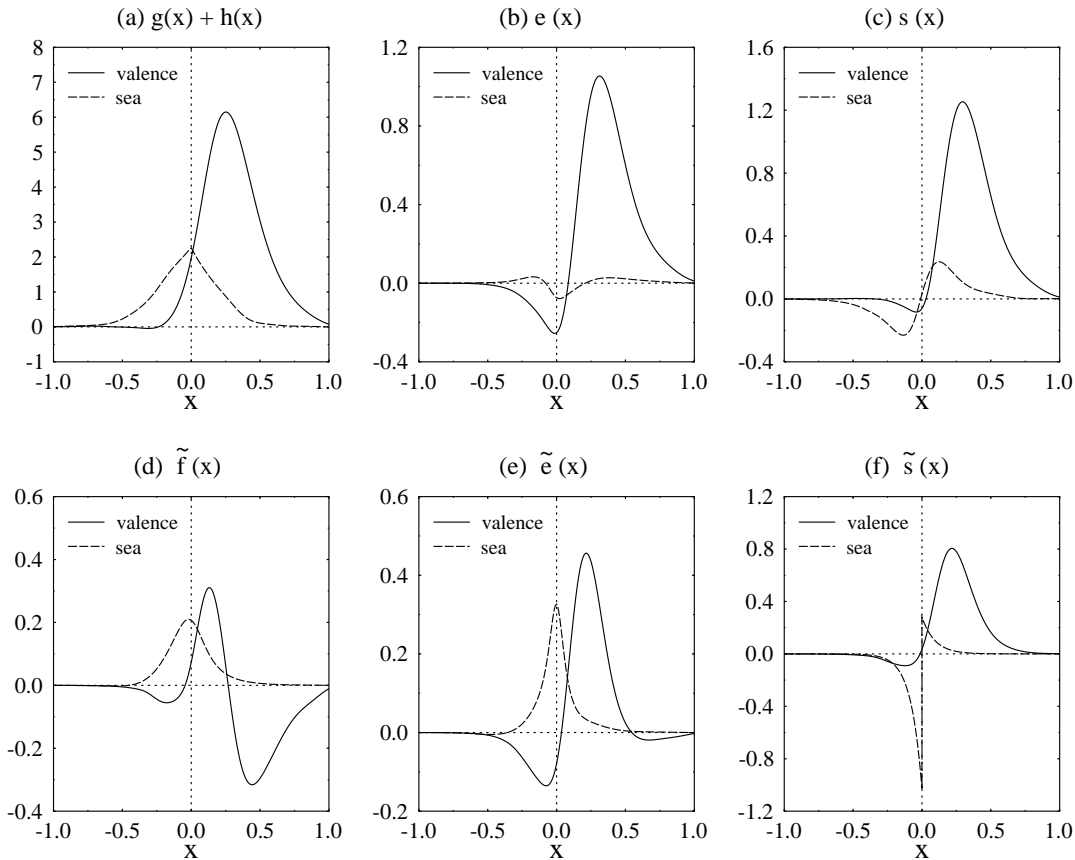}
\end{center}
\vspace*{-0.5cm}
\renewcommand{\baselinestretch}{1.20}
\caption{Six basic functions necessary for evaluating
longitudinally polarized distribution functions within the
SU(3) CQSM with $\Delta m_s$ corrections. The curves have the
same meanings as in Fig.~\ref{unpolmodel1}. \label{lgpolmodel1}}
\end{figure}

Next, in Fig.~\ref{lgpolmodel1}, we show six basic functions necessary
for evaluating longitudinally polarized distribution functions.
(Here, only the combination of $g(x)$ and $h(x)$ is shown, since it is
this combination that enters the theoretical expression of physical
distribution functions.) The first three functions $g(x) + h(x), e(x)$
and $s(x)$ appear in the leading $O(\Omega^0 + \Omega^1)$ terms, while
the remaining three functions $\tilde{f} (x), \tilde{e} (x)$ and
$\tilde{s} (x)$ are contained in the SU(3) symmetry breaking
corrections to the longitudinally polarized distribution functions.
One again sees that the contributions of Dirac-sea quarks have
appreciable effects on the total distributions, although they are less
significant than the case of the unpolarized distribution
functions \cite{SRW98}.
Among others, we point out that the function $g(x) + h(x)$ receives
appreciably large and positive vacuum polarization contributions in the
negative (as well as the positive) $x$ region.
We shall discuss later that this leads to a sizable large isospin
asymmetry for the longitudinally polarized sea-quark
(antiquark) distribution functions.

\begin{figure}[htb] \centering
\begin{center}
 \includegraphics[width=15.0cm]{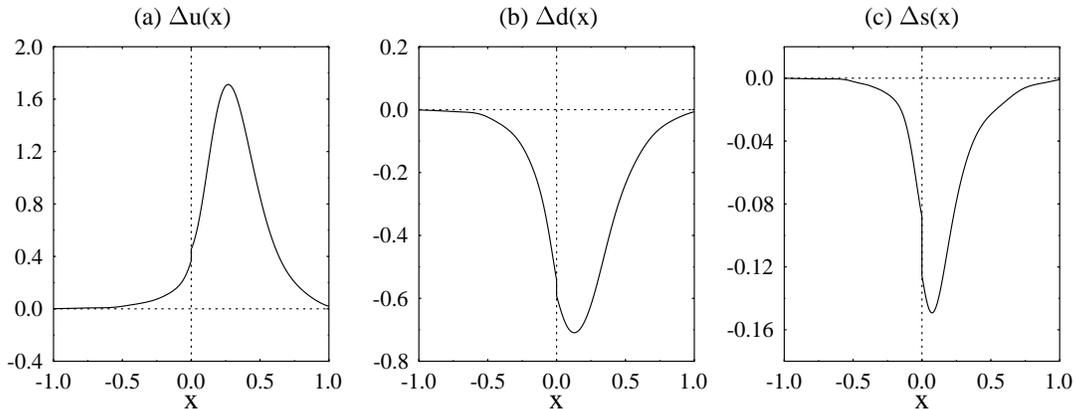}
\end{center}
\vspace*{-0.5cm}
\renewcommand{\baselinestretch}{1.20}
\caption{The longitudinally polarized distribution functions
with respective flavors. The distribution functions with negative
arguments should be interpreted as antiquark distributions
according to the rule $\Delta q(-x) = \Delta \bar{q}(x)$
with $0 < x <1$. \label{lgpolmodel2}}
\end{figure}

Using the above basic functions, we can calculate the longitudinally
polarized distribution functions with any flavors.
They are shown in Fig.~\ref{lgpolmodel2}.
Here, the polarized distributions
in the negative $x$ region should be interpreted as the polarized
antiquark ones according to the rule (5) of II, i.e.
$\Delta q(-x) = \Delta \bar{q} (x)$ with $0 < x < 1$.
From these figures, one can, for instance, read from these three
figures that $\Delta \bar{u} (x) > 0, \Delta \bar{d} (x) < 0$, and
$\Delta \bar{s} (x) < 0$. More detailed discussion of this interesting
predictions of the SU(3) CQSM will be given later.

\begin{figure}[htb] \centering
\begin{center}
 \includegraphics[width=13.0cm]{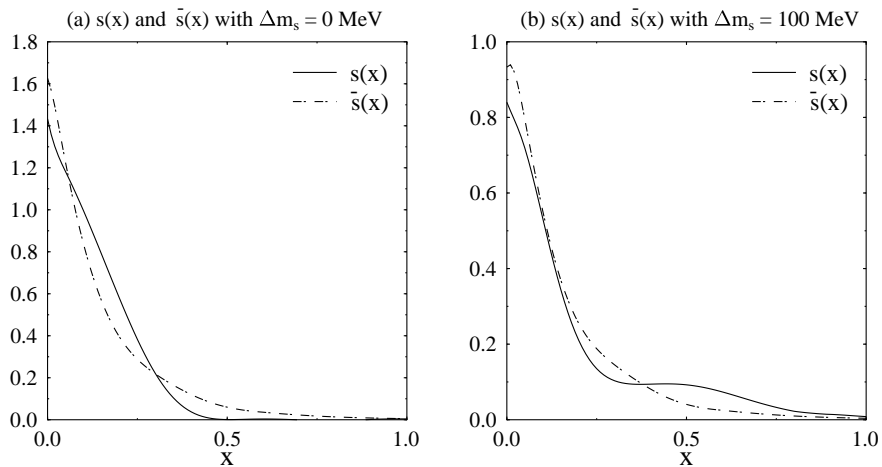}
\end{center}
\vspace*{-0.5cm}
\renewcommand{\baselinestretch}{1.20}
\caption{The theoretical predictions for the unpolarized $s$-
and $\bar{s}$-quark distributions at the model energy scale.
The left figure is obtained without $\Delta m_s$ corrections,
while the right one is with the value of
$\Delta m_s = 100 \,\mbox{MeV}$. \label{sdistunp}}
\end{figure}
\vspace{-5mm}
\begin{figure}[htb] \centering
\begin{center}
 \includegraphics[width=13.0cm]{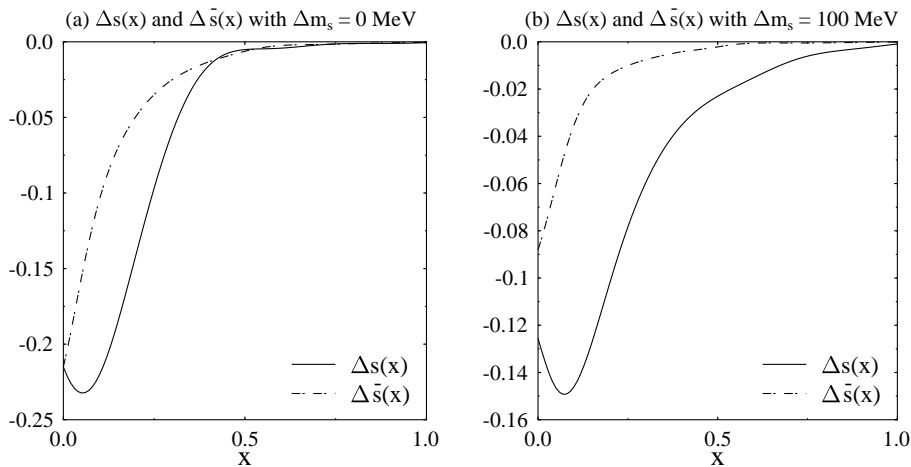}
\end{center}
\vspace*{-0.5cm}
\renewcommand{\baselinestretch}{1.20}
\caption{The theoretical predictions for the longitudinally
polarized $s$- and $\bar{s}$-quark distributions at the
model energy scale. The left figure shows the result obtained
without $\Delta m_s$ corrections, while the right one
corresponds to the result obtained with $\Delta m_s = 
100 \,\mbox{MeV}$. \label{sdistlgp}}
\end{figure}
\vspace{5mm}

Shown in Fig.~\ref{sdistunp} are the final predictions of
the SU(3) CQSM for the
unpolarized $s$- and $\bar{s}$-quark distributions at the 
model energy scale. The left panel shows the result obtained 
in the chiral limit, i.e. without SU(3) symmetry breaking effects, 
while the right panel corresponds to the result obtained after
introducing $\Delta m_s$ corrections.
One sees that the $s$-$\bar{s}$ asymmetry of the unpolarized 
distribution functions certainly exists.
The difference $s(x) - \bar{s} (x)$ has some oscillatory behavior
with several zeros as a function of $x$. This is of course due to 
the following two general constraints of the PDF :  
the positivity constraint for the unpolarized distributions 
and the strangeness quantum number conservations.
Comparing the two figures, one also finds that $s(x) - \bar{s} (x)$ 
is extremely sensitive to the SU(3) symmetry breaking effects.
Fig.~\ref{sdistlgp} shows the theoretical predictions for
the longitudinally 
polarized strange quark distributions. In the chiral limit case,
the $s$- and $\bar{s}$-quarks are both negatively polarized, although
the magnitude of $\Delta \bar{s} (x)$ is smaller than that of
$\Delta s(x)$.
After introducing $\Delta m_s$ corrections, $\Delta s(x)$ remains 
large and negative, while $\Delta \bar{s} (x)$ becomes very small 
although slightly negative.

\begin{figure}[htb] \centering
\begin{center}
 \includegraphics[width=14.0cm]{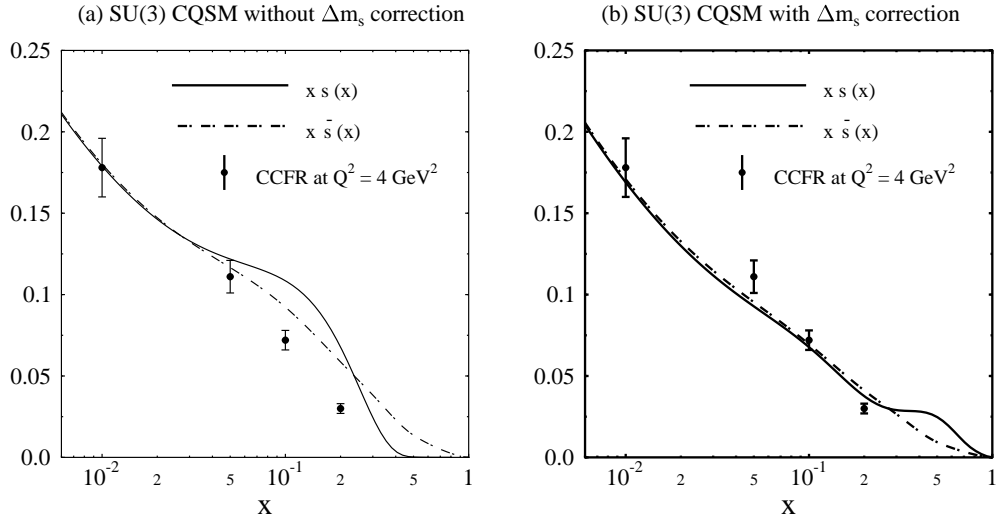}
\end{center}
\vspace*{-0.5cm}
\renewcommand{\baselinestretch}{1.20}
\caption{The theoretical unpolarized distribution functions
$s(x)$ and $\bar{s}(x)$ at $Q^2 = 4 \,\mbox{GeV}^2$ in
comparison with the corresponding CCFR data obtained under
the assumption $s(x) = \bar{s}(x)$ \cite{CCFR96}.
The left panel shows the
result obtained without $\Delta m_s$ corrections, whereas the
right panel represents the one with $\Delta m_s = 100 \,\mbox{MeV}^2$.
\label{ccfr04gev}}
\end{figure}

\begin{figure}[htb] \centering
\begin{center}
 \includegraphics[width=14.0cm]{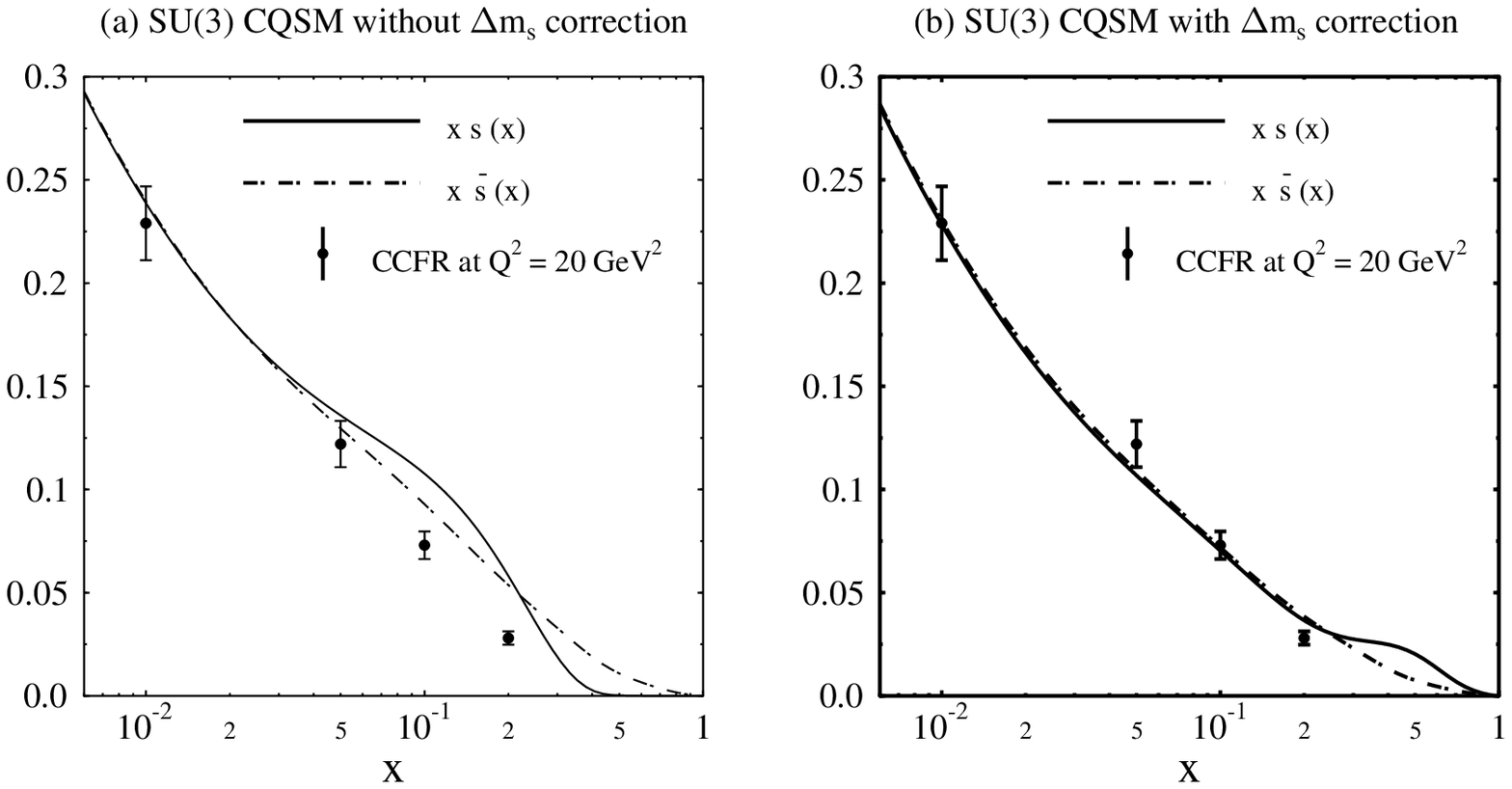}
\end{center}
\vspace*{-0.5cm}
\renewcommand{\baselinestretch}{1.20}
\caption{The theoretical unpolarized distribution functions $s(x)$
and $\bar{s}(x)$ at $Q^2 = 20 \,\mbox{GeV}^2$ in comparison
with the corresponding CCFR data \cite{CCFR96}.
The meanings of the curves
are the same as in Fig.~\ref{ccfr04gev} \label{ccfr20gev}}
\end{figure}

To sum up, it is a definite conclusion of our theoretical 
analysis that the particle-antiparticle asymmetry of the 
strange-quark excitation in the nucleon is most likely to exist.
Furthermore, the magnitude of the asymmetry seems more profound for 
the longitudinally polarized distribution than for the unpolarized 
one reflecting the fact that, for the polarized one, there exists no
conservation laws that prevents the generation 
of asymmetry. To understand the physical origin of these observations, 
it may be interesting to recall a simple argument of Brodsky and Ma
based on the light-cone meson baryon fluctuation model \cite{BM96}.
(See also Ref.~\cite{ST87}.) According to them, the intrinsic 
strangeness excitation in the proton is mainly due to the virtual 
$K^+ \Lambda$ dissociation process. Because of parity conservation, 
the relative orbital angular momentum of this two particle system 
must be odd, most probably be p-wave state. Using the Clebsh-Gordan
decomposition of this p-wave state,
\begin{eqnarray}
 | K^+ \Lambda (J = \frac{1}{2}, J_z = \frac{1}{2}) \rangle
 &=& \sqrt{\frac{3}{2}} \,| L = 1, L_z = 1 \rangle 
 | S = \frac{1}{2}, S_z = -\frac{1}{2} ) \nonumber \\  
 &-& \sqrt{\frac{1}{2}} \,| L = 1, L_z = 0 \rangle 
 | S = \frac{1}{2}, S_z = + \frac{1}{2} \rangle ,
\end{eqnarray}
one easily finds that the average spin projection of $\Lambda$ 
in the proton is negative. Because the $\Lambda$ spin mostly 
comes from the s-quark in it, it then immediately follows that 
the $s$-quark in the proton is negatively polarized. The situation
is entirely different for the $\bar{s}$-quark. Since the
$\bar{s}$-quark is contained in $K^+$ meson with zero spin, it
follows that the net spin of $\bar{s}$ in $K^+$ and consequently
in proton is zero. Note that whole these arguments are qualitatively
consistent with the predictions of the CQSM.
This indicates that the kaon cloud effects are automatically taken
into account by the collective rotation in the flavor SU(3) space,
a basic dynamical assumption of the SU(3) CQSM.

Now we want to make some preliminary comparisons with the existing
high-energy data for the strange quark distributions.
Fig.~\ref{ccfr04gev} and Fig.~\ref{ccfr20gev} respectively
shows the theoretical distributions
evolved to $Q^2 = 4 \mbox{GeV}^2$ and $Q^2 = 20 \mbox{GeV}^2$ in
comparison with the corresponding result of CCFR (NLO) analyses of
the neutrino-induced charm production carried out with the assumption
$\bar{s} (x) = s(x)$. In both figures, the solid and long-dashed 
curves represent the theoretical $s$- and $\bar{s}$-quark 
distributions, while the left panel shows the predictions 
obtained with $\Delta m_s = 0$ and the right panel shows those 
obtained with $\Delta m_s = 100 \mbox{MeV}$.
As was intuitively anticipated, the SU(3) 
symmetry breaking effects considerably suppress the magnitude 
of $s(x)$ and $\bar{s} (x)$ at the moderate range of $x$.   
The final theoretical predictions obtained with the $\Delta m_s$
corrections appears to be qualitatively consistent with the 
CCFR data, although various uncertainties of the phenomenological 
data for $s(x)$ and $\bar{s} (x)$ should not be forgotten.

\begin{figure}[htb] \centering
\begin{center}
 \includegraphics[width=9.0cm]{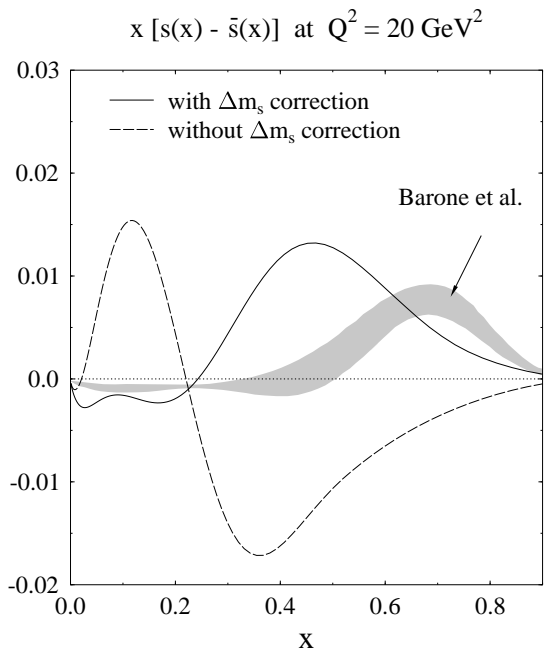}
\end{center}
\vspace*{-10mm}
\renewcommand{\baselinestretch}{1.20}
\caption{The theoretical predictions for the difference of
$s$- and $\bar{s}$-quark distributions at $Q^2 = 20 \,\mbox{GeV}^2$
in comparison with the corresponding result of Barone et al's
global analysis including neutrino data \cite{BPZ00}.
Here, the solid and dashed
curves respectively stand for the predictions of the SU(3) CQSM
with and without $\Delta m_s$ corrections. \label{diffssb}}
\end{figure}

Very recently, Barone et al. carried out quite elaborate 
global analysis of the DIS data, especially by using all the 
presently-available neutrino data also, and they obtained some 
interesting information even for the asymmetry of the $s$-
and $\bar{s}$-quark distributions \cite{BPZ00}.
Fig.~\ref{diffssb} shows
the comparison with their fit for the difference $s(x) - \bar{s}(x)$
at $Q^2 = 20 \mbox{GeV}^2$. Here, the thin shaded area represent the 
phenomenologically favorable region for this difference function
obtained by Barone et al's global fit. On the other hand, 
the solid and dashed curves are the predictions of the 
SU(3) CQSM, respectively obtained with and without the $\Delta m_s$
correction. One sees that, the difference $s(x) - \bar{s} (x)$
is extremely sensitive to the SU(3) symmetry breaking effects 
and that, after inclusion of it, the theory reproduces the 
qualitative tendency of the phenomenologically obtained behavior 
of $s(x) - \bar{s} (x)$ although not perfectly.

\begin{figure}[htb] \centering
\begin{center}
 \includegraphics[width=9.0cm]{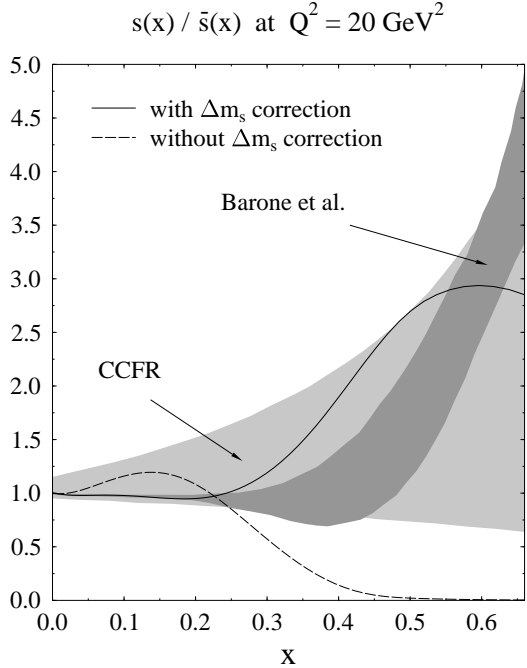}
\end{center}
\vspace*{-10mm}
\renewcommand{\baselinestretch}{1.20}
\caption{The theoretical predictions for the ratio of
$s$- and $\bar{s}$-quark distributions at $Q^2 = 20 \,\mbox{GeV}^2$
in comparison with the results of CCFR analysis \cite{CCFR96} and
of Barone et al's global fit \cite{BPZ00}.
The meaning of the curves are the same as in
Fig.~\ref{diffssb} \label{ratssb}}
\end{figure}

This tendency is more clearly seen in the ratio of $s(x)$ and 
$\bar{s} (x)$ at $Q^2 = 20 \mbox{GeV}^2$. In Fig.~\ref{ratssb},
the solid and dashed 
curves are the predictions of the SU(3) CQSM with and without 
the $\Delta m_s$ corrections, while the thin and thick shaded 
areas represent the phenomenologically favorable regions for 
this ratio, respectively obtained by the CCFR group and by Barone 
et al. One clearly sees that the observed tendency of this ratio 
is reproduced (at least qualitatively) only after including
the SU(3) symmetry breaking effects.

Now, turning to the spin-dependent distribution functions, the 
quality of the presently available semi-inclusive data is 
rather poor, so that the analyses are mainly limited to the 
inclusive DIS data alone. (There exist some combined
analyses of inclusive and semi-inclusive polarized DIS
data \cite{MY00},\cite{FS00},\cite{BB02}.)
This forces them to introduce several 
simplifying assumptions in the fittings. For instance, many 
previous analyses have used the apparently groundless assumption 
of a flavor-symmetric polarized sea, i.e. 
$\Delta \bar{u} (x) = \Delta  \bar{d} (x) = \Delta \bar{s} (x)$
\cite{GS96}.
Another analysis assumed that $\Delta q_3 (x) = c \,\Delta q_8 (x)$
with $c$ being a constant. Probably, the most ambitious analyses
free from these {\it ad hoc} assumptions on the 
sea-quark distributions are those of Leader, Sidorov and Stamenov
(LSS) \cite{LSS00}. (See also Ref.~\cite{GRDV01}.)
We recall that they also investigated the sensitivity of their 
fit to the size of the SU(3) symmetry breaking effect.
(Although they did not take account of the possibility that 
$\Delta s(x) \neq \Delta \bar{s} (x)$, this simplification is 
harmless, because only the combination $\Delta s(x) 
+ \Delta \bar{s} (x)$ appears in their analyses of DIS data.)

To compare the theoretical distributions of the SU(3) CQSM 
with the LSS fits given at $Q^2 = 1 \mbox{GeV}^2$, we must consider 
the fact that their analyses are carried out in the so-called 
JET scheme (or the chirally invariant factorization scheme). 
To take account of this, we start with the theoretical 
distribution functions $\Delta u (x), \Delta \bar{u} (x), 
\Delta d (x), \Delta \bar{d} (x), \Delta s (x)$ and 
$\Delta \bar{s} (x)$, which are taken as the initial scale 
distribution functions given at $Q^2_{ini} = 0.30 \mbox{GeV}^2$.    
Under the assumption that $\Delta g(x) = 0$ at this initial energy 
scale, we solve the DGLAP equation in the standard $\overline{MS}$ 
scheme with the gauge-invariant factorization scheme to obtain 
the distributions at $Q^2 = 1 \mbox{GeV}^2$.
The corresponding distribution
functions in the JET scheme are then obtained by the transformation
\begin{eqnarray}
 \Delta \Sigma (x.Q^2)_{JET} &=& 
 \Delta \Sigma (x.Q^2)_{\overline{MS}}
 + \frac{\alpha_s (Q^2)}{\pi} N_f (1 - x) 
 \otimes \Delta g (x. Q^2)_{\overline{MS}} , \\ 
 \Delta g (x. Q^2)_{JET} &=& \Delta g (x.Q^2)_{\bar{MS}} ,
\end{eqnarray}
with $\Delta \Sigma (x.Q^2) = \sum^{N_f}_{i = 1} (\Delta q_i (x. Q^2)
+ \Delta \bar{q}_i (x. Q^2))$ being the flavor-singlet quark 
polarization. 

\begin{figure}[htb] \centering
\begin{center}
 \includegraphics[width=14.0cm]{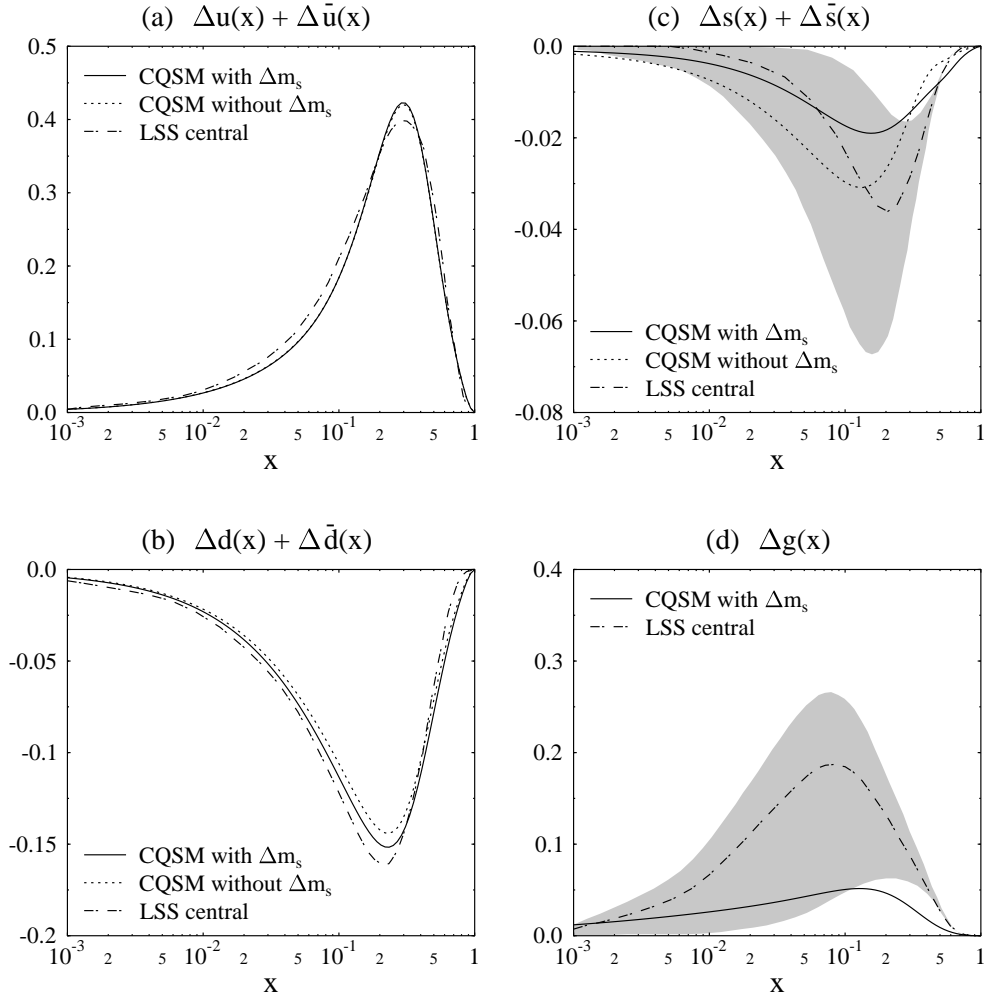}
\end{center}
\vspace*{-0.5cm}
\renewcommand{\baselinestretch}{1.20}
\caption{The theoretical distribution functions
(a) $x (\Delta u(x) + \Delta \bar{u}(x))$, 
(b) $x (\Delta d(x) + \Delta \bar{d}(x))$, 
(b) $x (\Delta s(x) + \Delta \bar{s}(x))$, and
(d) $x g(x)$ at $Q^2 = 1 \,\mbox{GeV}^2$, in comparison with
the corresponding LSS fits in the JET scheme \cite{LSS00}.
Here, the solid and dotted curves are the predictions of the
SU(3) CQSM with and without $\Delta m_s$ corrections, while the
central fit by LSS analyses are represented by the dash-dotted
curves. The large uncertainties for the strange-quark distribution
as well as the gluon distribution in the LSS fits are illustrated
by the shaded areas. \label{lsscomp}}
\end{figure}

Now we show in Fig.~\ref{lsscomp} the theoretical distributions 
$x (\Delta u (x) + \Delta \bar{u}(x)), 
x(\Delta d (x) + \Delta \bar{d} (x))$,
$x(\Delta s (x) + \Delta \bar{s} (x))$ and 
$\Delta g(x)$ at $Q^2 = 1 \mbox{GeV}^2$ in comparison with the 
corresponding LSS fits. The solid and dashed curves in these 
four figures are respectively the predictions of the SU(3) CQSM 
obtained with and without the $\Delta m_s$ corrections. 
To estimate the sensitivity of the fit to the SU(3) symmetry 
breaking effects, Leader et al. performed their fit by varying 
the value of axial charge $a_8$ from its SU(3) symmetric value 
$0.58$ within the range $0.40 \leq a_8 \leq 0.86$.
They found that the value of $\chi^2$-fit to the presently 
available DIS data are practically insensitive to the variation 
of  $a_8$, which in turn means that $a_8$ cannot be determined 
from the existing DIS data. Consequently, the distributions
$x [ \Delta u (x) + \Delta \bar{u} (x) ]$ and 
$x [ \Delta d (x) + \Delta \bar{d} (x) ]$
are insensitive to the SU(3) symmetry breaking effects and can 
be determined with little uncertainties. This is also confirmed by 
our theoretical analysis. In Fig.~\ref{lsscomp}(a) and
Fig.~\ref{lsscomp}(b), the solid 
and dashed curves are the predictions of the SU(3) CQSM for the 
distributions $x [ \Delta u (x) + \Delta \bar{u} (x) ]$ and
$x [ \Delta d (x) + \Delta \bar{d} (x) ]$ respectively obtained 
with and without the $\Delta m_s$ corrections. One confirms that 
they are nearly degenerate and that they reproduce well the results
of LSS fit. On the other hand, the distributions of the strange 
quarks and the gluons are very sensitive to the variation of 
the axial charge $a_8$, so that it brings about large uncertainties 
for these distributions in the LSS fit as illustrated by the 
shaded regions in Fig.~\ref{lsscomp}(c) and Fig.~\ref{lsscomp}(d).
The feature is again 
consistent with our theoretical analysis at least for the 
polarized strange-quark distributions. In fact, the SU(3) CQSM 
predicts that $x [ \Delta s (x) + \Delta \bar{s} (x) ]$ is large
and negative but the $\Delta m_s$ correction reduces its 
magnitude by a factor of about $0.6$.

\begin{figure}[htb] \centering
\begin{center}
 \includegraphics[width=10.0cm]{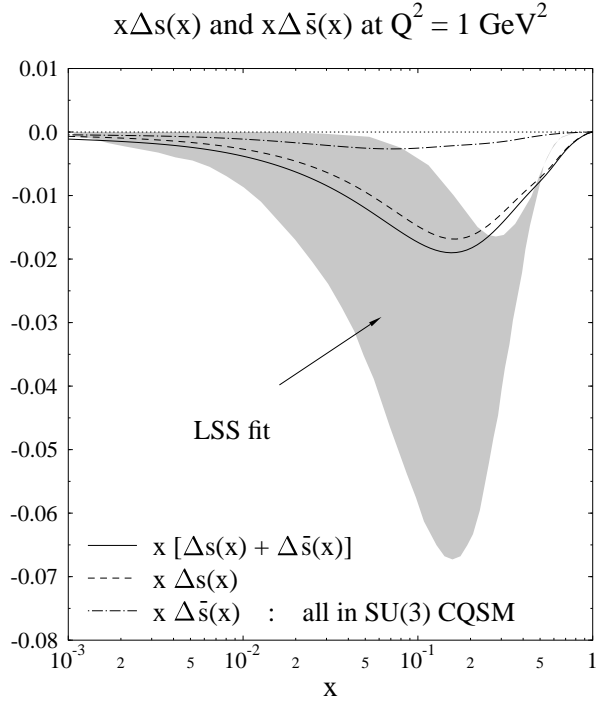}
\end{center}
\vspace*{-1.0cm}
\renewcommand{\baselinestretch}{1.20}
\caption{The theoretical prediction of the SU(3) CQSM for the
separate contributions of $s$- and $\bar{s}$-quarks to the
longitudinally polarized distribution functions
$x [\Delta s(x) + \Delta \bar{s}(x)]$ in comparison with
the LSS fit \cite{LSS00}. \label{delssbar}}
\end{figure}

A noteworthy feature of the theoretical predictions of the 
SU(3) CQSM is that the negative polarization of strange sea 
comes almost solely from $\Delta s (x)$, while 
$\Delta \bar{s} (x)$ is nearly zero.    
This is illustrated in Fig.~\ref{delssbar}.
The predicted sizable particle-antiparticle asymmetry of the 
polarized strange sea can be verified only by the near 
future experiments beyond the totally inclusive DIS scatterings 
such as the semi-inclusive DIS processes, the neutrino 
reactions, etc.

\begin{figure}[htb] \centering
\begin{center}
 \includegraphics[width=15.0cm]{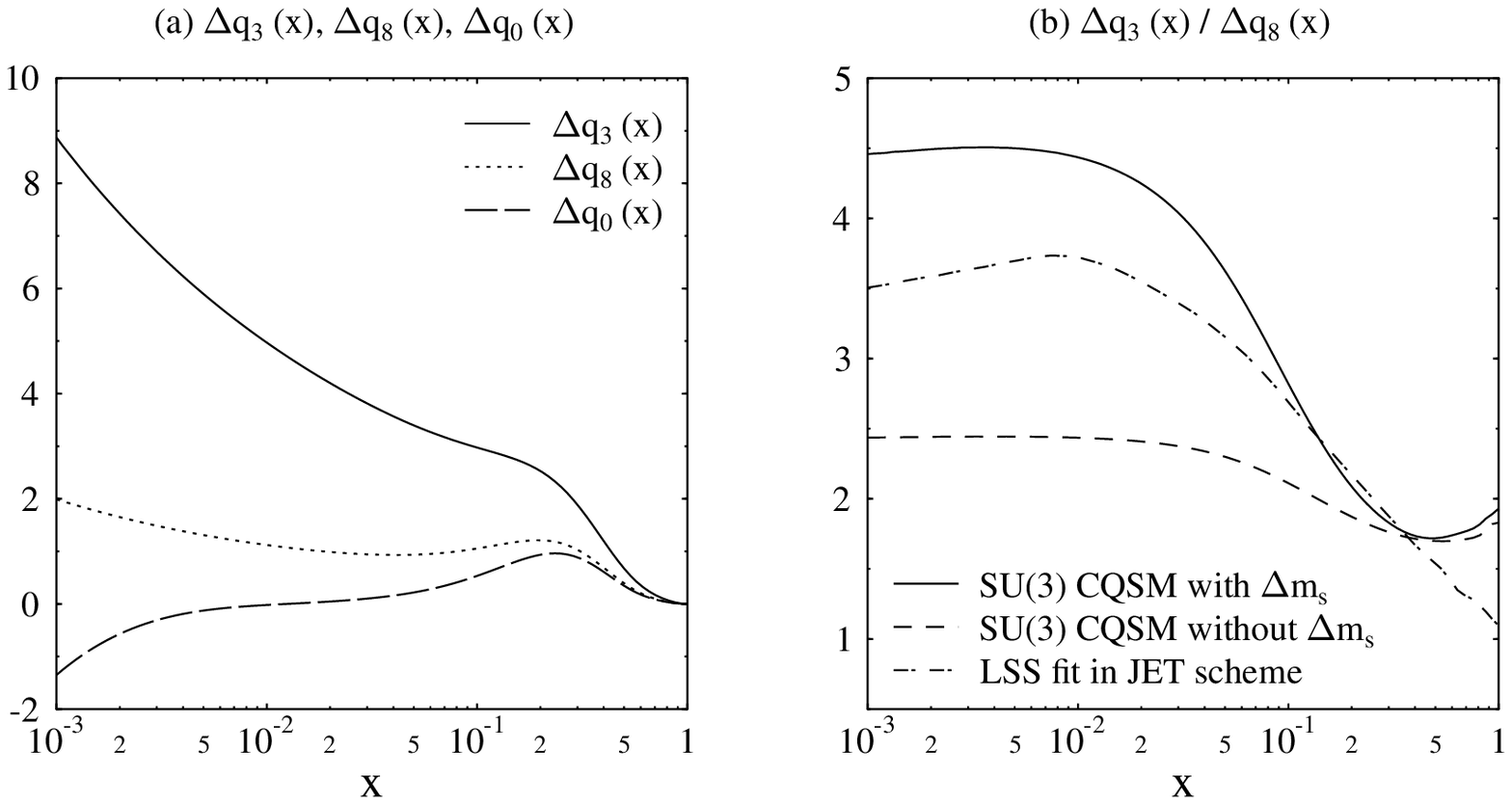}
\end{center}
\vspace*{-0.5cm}
\renewcommand{\baselinestretch}{1.20}
\caption{The flavor nonsinglet and single combinations of the
longitudinally polarized distribution functions and their ratio
at $Q^2 = 1 \,\mbox{GeV}^2$ in the JET scheme :
(a) $\Delta q_3 (x), \Delta q_8 (x)$, and $\Delta q_0 (x)$,
and (b) $\Delta q_3 (x) / \Delta q_8 (x)$.
In the right panel, the solid and dashed curves
are the predictions of the SU(3) CQSM with and without $\Delta m_s$
corrections, whereas the dash-dotted curve stands for the
corresponding LSS fit.
\label{delq3q8}}
\end{figure}

From the theoretical viewpoint, it is interesting to see the 
particular linear combinations of the distributions $\Delta u (x) 
+ \Delta \bar{u} (x), \Delta d (x) + \Delta \bar{d} (x)$ and
$\Delta s (x) + \Delta \bar{s} (x)$ given by
\begin{eqnarray}
 \Delta q_0 (x) &=& \Delta u (x) + \Delta \bar{u} (x) 
 + \Delta d (x) + \Delta \bar{d} (x)
 + \Delta s (x) + \Delta \bar{s} (x) , \\
 \Delta q_3 (x) &=& \Delta u (x) + \Delta \bar{u} (x)
 - \Delta d (x) - \Delta \bar{d} (x) , \\
 \Delta q_8 (x) &=& \Delta u (x) + \Delta \bar{u} (x)
 + \Delta d (x) + \Delta \bar{d} (x) 
 - 2 \,[ \Delta s (x) + \Delta \bar{s} (x) ] .
\end{eqnarray}
We show in Fig.~\ref{delq3q8}(a) the predictions of
the SU(3) CQSM for these densities at $Q^2 = 1 \mbox{GeV}^2$
in the JET scheme.
One clearly sees that $\Delta q_0 (x)$ is negative in the 
smaller $x$ region. One can also convince that the polarized 
strange quark densities given by
\begin{equation}
 \Delta s (x) + \Delta \bar{s} (x) = \frac{1}{3} \,
 [ \Delta q_0 (x) - \Delta q_8 (x) ] ,
\end{equation}
is certainly negative for all range of $x$.
Of special interest here is the difference or the ratio of 
$\Delta q_3 (x)$ and $\Delta q_8 (x)$, since, as already pointed 
out, some previous phenomenological analyses assume 
$\Delta q_3 (x) / \Delta q_8 (x) =$ constant with no justification.
The solid and dashed curves in Fig.~\ref{delq3q8}(b) are the
predictions of the SU(3) CQSM for this ratio, respectively obtained
with and without $\Delta m_s$ corrections, while dash-dotted curves 
is the corresponding result of LSS fit. After including the SU(3) 
symmetry breaking effects, one can say that the theory reproduces
the qualitative behavior of the LSS fit for this ratio.

Through the analyses so far, we have shown that the flavor SU(3) 
CQSM can give unique and interesting predictions for both of the
unpolarized and the longitudinally polarized strange 
quark distributions in the nucleon, all of which seems to be 
qualitatively consistent with the existing phenomenological 
information for strange quark distributions. A natural question
here is whether or not it is realistic enough as the flavor 
SU(2) CQSM has been  proved so. (We recall that the SU(2) CQSM 
reproduces almost all the qualitatively noticeable features 
of the presently available DIS data.) 
Which is more realistic model of the nucleon, the SU(2) CQSM 
or the SU(3) one?
Naturally, at least concerning one particular aspect, i.e. 
the problem of hidden strange-quark excitations in the nucleon, 
the SU(3) model is superior to the SU(2) model, since the
strange quark excitations in the nucleon can be treated only in
the former model. The question is then reduced to which model
gives more realistic descriptions for the $u,d$-flavor dominated
observables, which have been the objects of studies of the SU(2)
CQSM. To answer this question, we try to reanalyze several
interesting observables, which we have investigated before in the
SU(2) CQSM, here within the framework of the flavor SU(3) CQSM.

\begin{figure}[htb] \centering
\begin{center}
 \includegraphics[width=14.0cm]{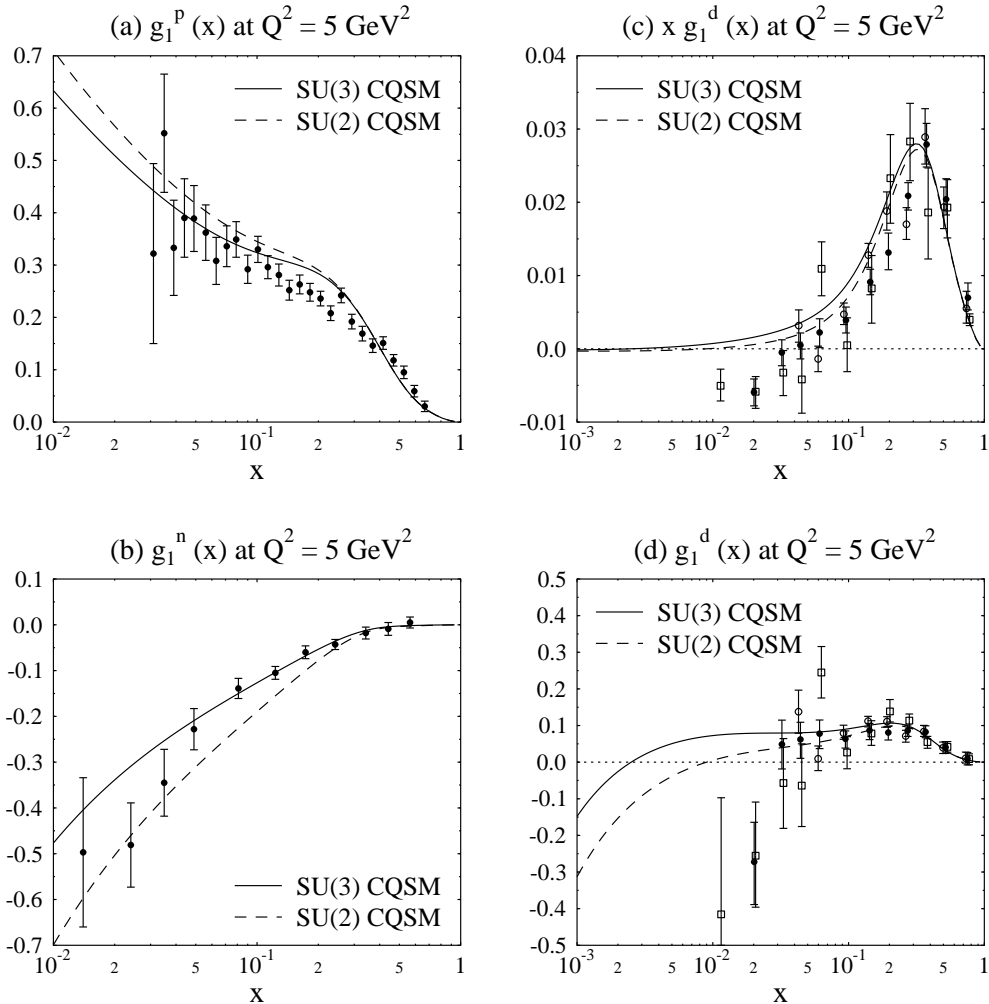}
\end{center}
\vspace*{-0.5cm}
\renewcommand{\baselinestretch}{1.20}
\caption{The theoretical predictions for the proton, neutron,
and deuteron spin structure functions $g_1^p (x), g_1^n (x)$, and
$g_1^d (x)$ at $Q^2 = 5 \,\mbox{GeV}^2$ in comparison with the
corresponding SLAC and SMC data. The solid and dashed curves
in these figures respectively stand for the predictions of
the SU(3) CQSM and those of the SU(2) CQSM.
The black circles in (a) and (b) are the E143 \cite{E143} and
the E154 data \cite{E154}, while the diamonds, the circles and
the squares in (c) represent the E143 \cite{E143}, the E155
\cite{E155}and the SMC data \cite{SMC98}. 
\label{emc}}
\end{figure}

At this opportunity, the calculation in the SU(2) model were 
redone, because there is a little change in the theoretical 
treatment of the $O(\Omega^1)$ contribution to the 
longitudinally polarized distribution functions as was explained 
in the previous section. 
First, in Fig.~\ref{emc}, we show the theoretical predictions of
the SU(3) and SU(2) CQSM for the longitudinally polarized structure 
functions of the proton, the neutron and the deuteron in 
comparison with the corresponding EMC and SMC data at 
$Q^2 = 5 \mbox{GeV}^2$. Here, the solid and dashed curves are the 
predictions of the SU(3) and SU(2) CQSM, respectively.
The black circles in Fig.~\ref{emc}(a) are E143 data, whereas those 
in Fig.~\ref{emc}(b) are E154 data \cite{E154}.
On the other hand, the black circles,
white circles and white squares in Fig.~\ref{emc}(c) and
Fig.~\ref{emc}(d) correspond to the E143 \cite{E143},
E155 \cite{E155} and SMC data \cite{SMC98},
respectively.
Comparing the predictions of the two versions of the CQSM,
one notices two features. First, the magnitudes of $g_1^p (x)$
and $g_1^n (x)$ are reduced a little when going from the SU(2)
model to the SU(3) one. As we will discuss shortly, this
feature can be understood as a reduction of the isovector
axial charge in the SU(3) CQSM. Another feature is that the
small $x$ behavior of the deuteron structure function
(the flavor singlet one) becomes slightly worse in the SU(3)
model. (This is due to the SU(3) symmetry breaking effects.)
These differences between the predictions of the SU(3) CQSM and
the SU(2) one are very small, however. Considering the qualitative
nature of our model as an effective low energy theory of QCD,
we may be allowed to say that both models reproduces the
experimental data fairly well.

\begin{table}[htb]
\begin{center}
\renewcommand{\baselinestretch}{1.20}
\newcommand{\lw}[1]{\smash{\lower2.ex\hbox{#1}}}
\caption{The predictions of the SU(3) and SU(2) CQSM for the
axial charges, the quark polarization
$\Delta q \equiv \int_0^1 \,[\Delta q(x) + \Delta \bar{q}(x)] \,dx$
of each flavor, and the basic coupling constant of SU(3),
in comparison with phenomenological information. Here, the
experimental values for $g_A^{(3)}, g_A^{(8)}, F, D$ and $F/D$ are
from \cite{CR93}, while $\Delta u, \Delta d, \Delta s$ and
$g_A^{(0)}$ corresponds to the values at $Q^2 = 10 \,\mbox{GeV}^2$
given in \cite{EK95}. \label{acharge}}
\vspace{15mm}
\renewcommand{\arraystretch}{1.0}
\begin{tabular}{|c|c|c|c|} \hline
 \, & \ SU(2) CQSM \ & \ SU(3) CQSM \ & \ \ \ Experiment \ 
\ \ 
 \\ \hline\hline
 \ $g_A^{(3)}$ \ & 1.41 & 1.20 & 1.257 $\pm$ 0.016 \\ \hline
 \ $g_A^{(8)}$ \ & --- & 0.59 & 0.579 $\pm$ 0.031 \\ \hline
 \ $g_A^{(0)}$ \ & 0.35 & 0.36 & 0.31 $\pm$ 0.07 \\ \hline\hline
 \ $\Delta u$ \ & 0.88 & 0.82 & 0.82 $\pm$ 0.03 \\ \hline
 \ $\Delta d$ \ & -0.53 & -0.38 & -0.44 $\pm$ 0.03 \\ \hline
 \ $\Delta s$ \ & 0 & -0.08 & -0.11 $\pm$ 0.03 \\ \hline\hline
 \ $F$ \ & --- & 0.45 & 0.459 $\pm$ 0.008 \\ \hline
 \ $D$ \ & --- & 0.76 & 0.798 $\pm$ 0.008 \\ \hline
 \ $F / D$ \ & ---
 & 0.59 & 0.575 $\pm$ 0.016 \\ \hline
\end{tabular}
\end{center}
\renewcommand{\baselinestretch}{1.20}
\end{table}

As mentioned above, the reduction of the magnitudes of
$g_1^p (x)$ and $g_1^n (x)$ in the SU(3) CQSM can be traced back to
the change of the isovector charge, which is related to the first
moment of $g_1^p (x) - g_1^n (x)$. We show in Table 1 the predictions
of the SU(3) and SU(2) CQSM for the flavor nonsinglet as well as
flavor singlet axial charges, the quark polarization
of each flavor defined as $\Delta q = \int_0^1 \,
[\Delta q(x) + \Delta \bar{q}(x) ]\, dx$,
in comparison with some phenomenological
information. One sees that, aside from the addition of
the strange quark degrees of freedom, a main change when going
from the SU(3) model to the SU(2) one is a decrease of isovector
axial charge $g_A^{(3)}$, while the flavor singlet axial charge
$g_A^{(0)}$ is almost unchanged. Corresponding to this\
reduction of $g_A^{(3)}$, the magnitudes of $\Delta u$ and
$\Delta d$ are both reduced a little. 
Also shown in this table is the fundamental coupling constants
$F$ and $G$ in the flavor SU(3) scheme as well as their ratio.
They are all qualitatively consistent with the phenomenological
information. Interestingly, the predicted ratio $F / D$ is
very close to that of the naive SU(6) model, i.e. $3/5$,
even though such dynamical symmetry is not
far from being justified in our theoretical framework.

\begin{figure}[htb] \centering
\begin{center}
 \includegraphics[width=15.0cm]{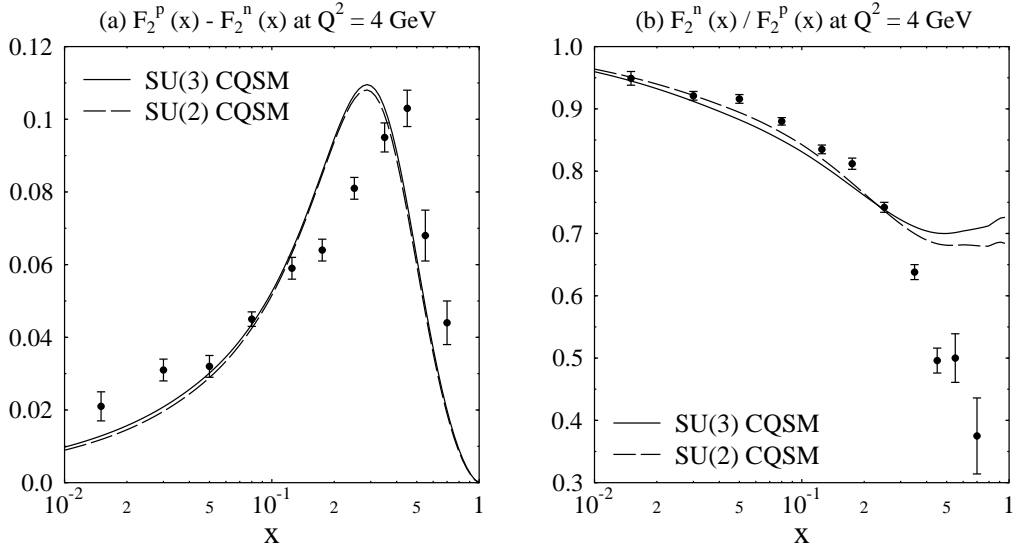}
\end{center}
\vspace*{-0.5cm}
\renewcommand{\baselinestretch}{1.20}
\caption{The theoretical predictions of the two versions of the
CQSM for $F_2^p (x) - F_2^n (x)$ and $F_2^n (x) / F_2^p (x)$
at $Q^2 = 4 \,\mbox{GeV}^2$ are compared with the NMC data given
at the corresponding energy scale \cite{NMC91}. \label{nmc}}
\end{figure}

Next, we go back to the spin-independent observables.
The solid and dashed curves in Fig.~\ref{nmc}(a) stand for the
predictions of the SU(3) and SU(2) CQSM for the difference
$F^p_2 (x) - F^n_2 (x)$ of the proton and neutron structure function 
$F_2 (x)$, in comparison with the corresponding NMC data 
at $Q^2 = 4 \mbox{GeV}^2$. One the other hand, the solid and dashed 
curves in Fig.~\ref{nmc}(b) are the predictions of the SU(3) and SU(2) 
CQSM for the ratio $F^n_2 (x) / F^p_2 (x)$ in comparison with 
the NMC data. One confirms that these difference and the ratio 
functions are rather insensitive to the flavor SU(3) generalization 
of the model and that the success of the SU(2) CQSM is basically 
taken over by the SU(3) model.

\begin{figure}[htb] \centering
\begin{center}
 \includegraphics[width=10.0cm]{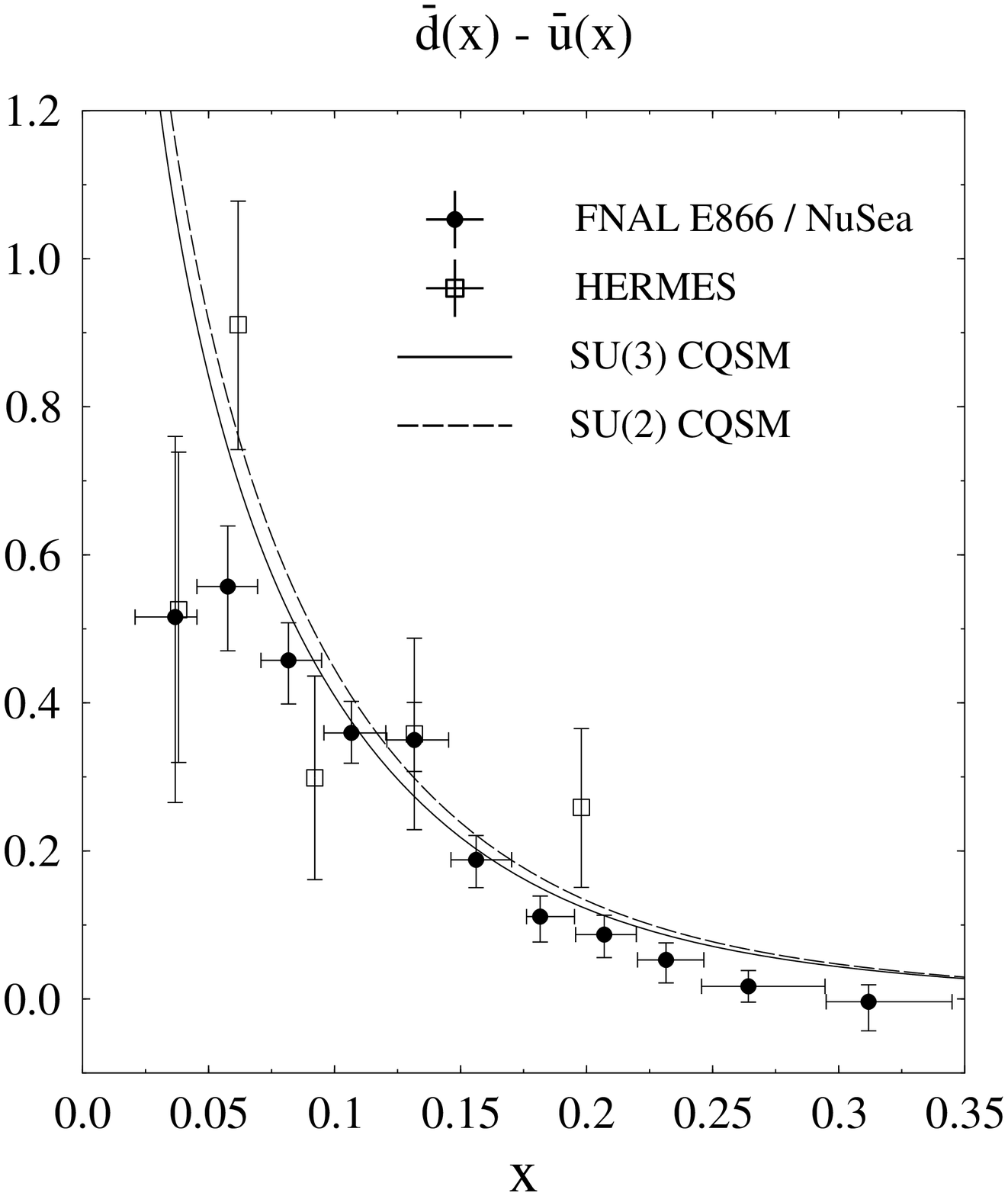}
\end{center}
\vspace*{-0.5cm}
\renewcommand{\baselinestretch}{1.20}
\caption{The theoretical predictions of the SU(3) and SU(2) CQSM
for the unpolarized antiquark distribution $\bar{d}(x) - \bar{u}(x)$
at $Q^2 = 54 \,\mbox{GeV}^2$ in comparison with the HERMES
\cite{HERMES} and E866 data \cite{E866}. \label{nusea}}
\end{figure}

In Fig.~\ref{nusea}, the theoretical 
predictions of both models for the sea-quark distribution 
$\bar{d} (x) - \bar{u} (x)$ is compared with the corresponding 
E866 data at $Q^2 = 54 \mbox{GeV}^2$ \cite{E866} and with HERMES
data at $Q^2 = 4 \mbox{GeV}^2$ \cite{HERMES}, for reference.
The isospin asymmetry of the 
sea-quark distributions or the magnitude of 
$\bar{d} (x) - \bar{u} (x)$ turns out to become a little 
smaller in the SU(3) model than in the SU(2) model, although 
the change is fairly small.

\begin{figure}[htb] \centering
\begin{center}
 \includegraphics[width=10.0cm]{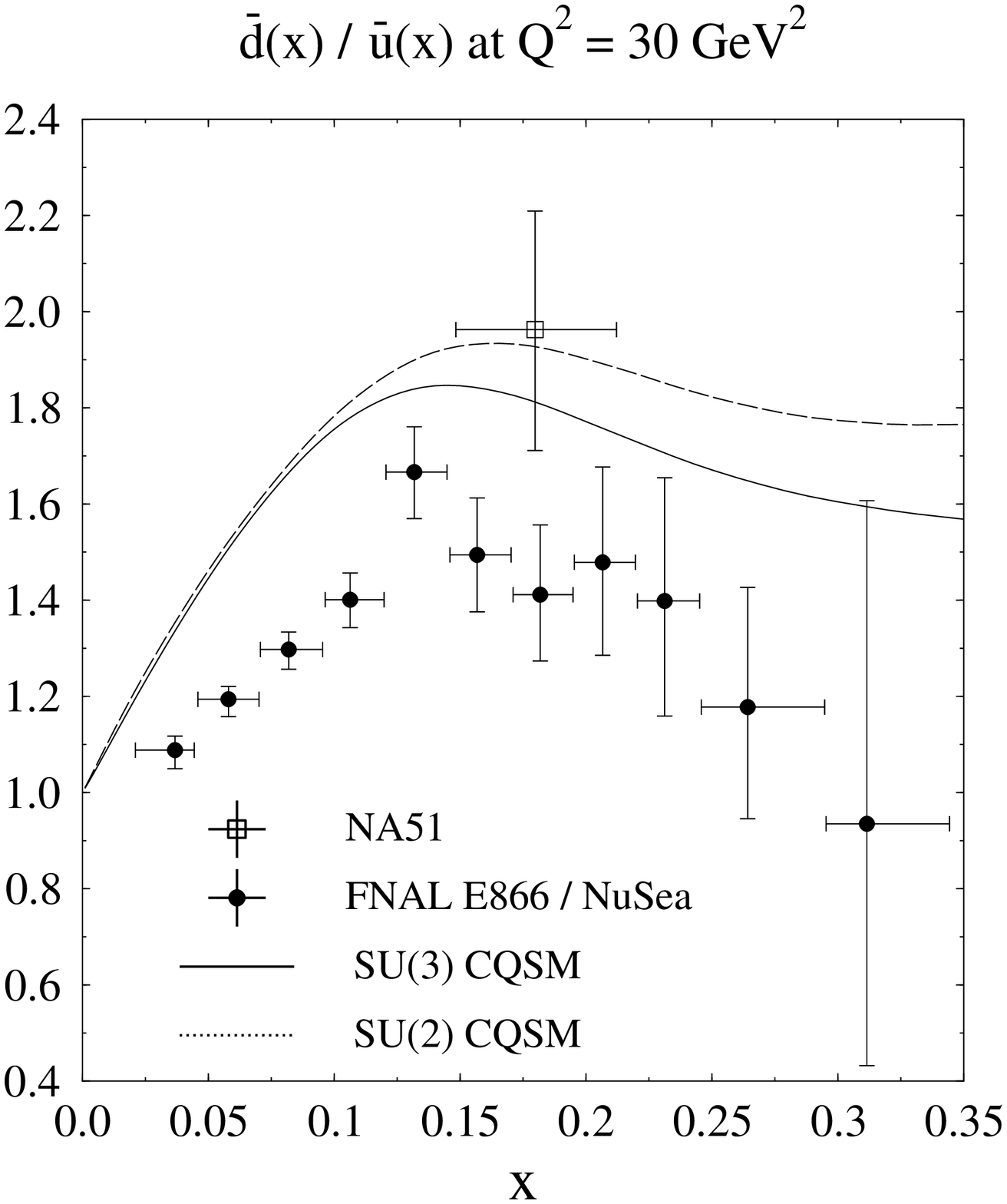}
\end{center}
\vspace*{-0.5cm}
\renewcommand{\baselinestretch}{1.20}
\caption{The theoretical predictions of the SU(3) and SU(2) CQSM
for the ratio $\bar{d}(x) / \bar{u}(x)$ in the proton as a function
of $x$ in comparison with the result of E866 analysis \cite{E866}.
Also shown is the result from NA51 \cite{NA51}, plotted as an open box.
\label{na51}}
\end{figure}

\begin{figure}[htb] \centering
\begin{center}
 \includegraphics[width=14.0cm]{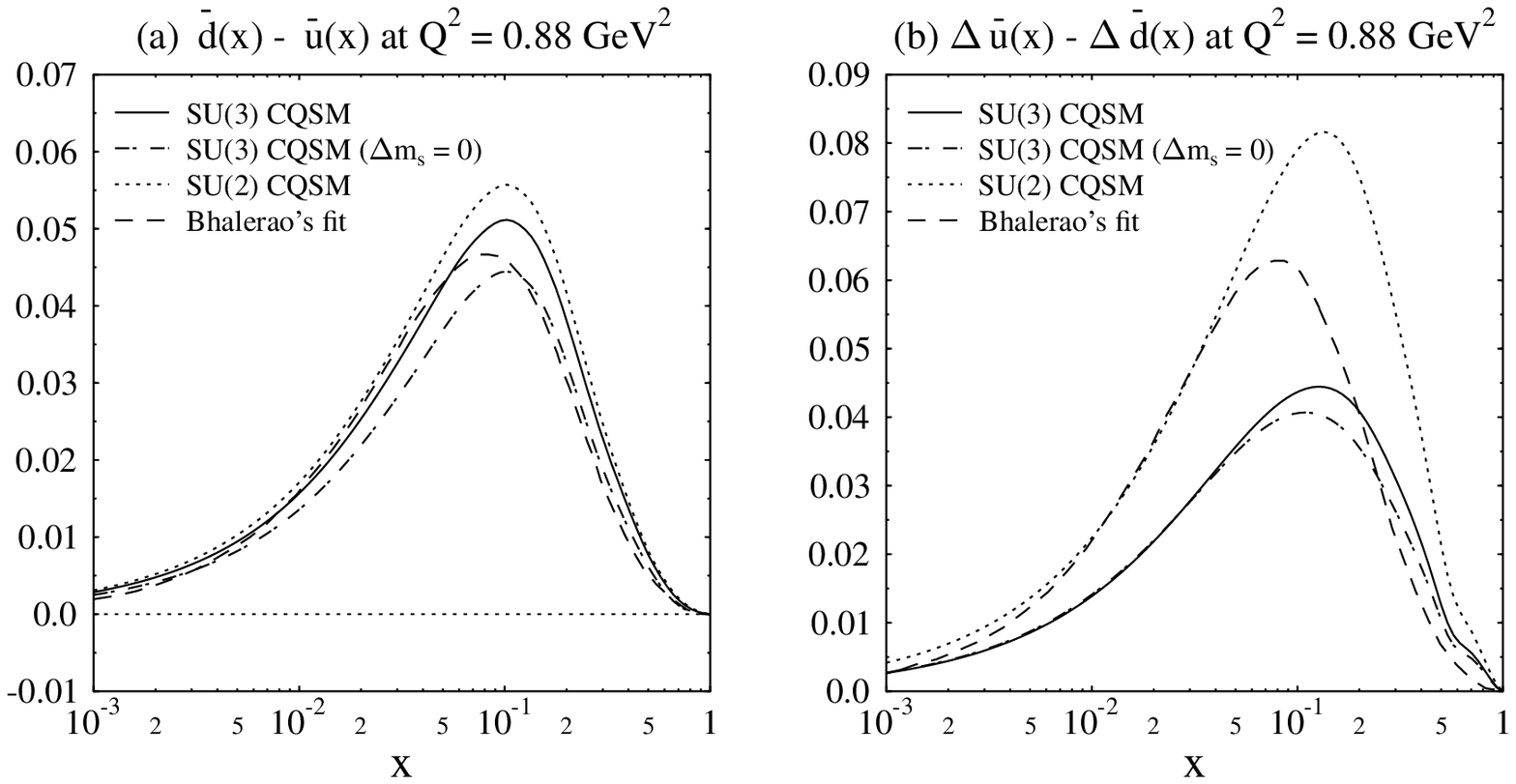}
\end{center}
\vspace*{-0.5cm}
\renewcommand{\baselinestretch}{1.20}
\caption{The isospin asymmetries of the light-flavor
sea-quark distribution functions evaluated at
$Q^2 = 0.88 \,\mbox{GeV}^2$ in the standard $\overline{MS}$
factorization scheme with gauge invariant regularization.
The left figure shows the unpolarized distribution
$x [\bar{d}(x) - \bar{u}(x)]$, while the right figure
represents the longitudinally polarized one
$x [\Delta \bar{u}(x) - \Delta \bar{d}(x) ]$.
In both figures, the solid and dash-dotted curves are the
predictions of the SU(3) CQSM with and without $\Delta m_s$
corrections, whereas the dotted curves are those of the SU(2) CQSM.
Bhalerao's semi-theoretical predictions are also shown for
comparison \cite{B01}. \label{asymsea}}
\end{figure}

\begin{figure}[htb] \centering
\begin{center}
 \includegraphics[width=14.0cm]{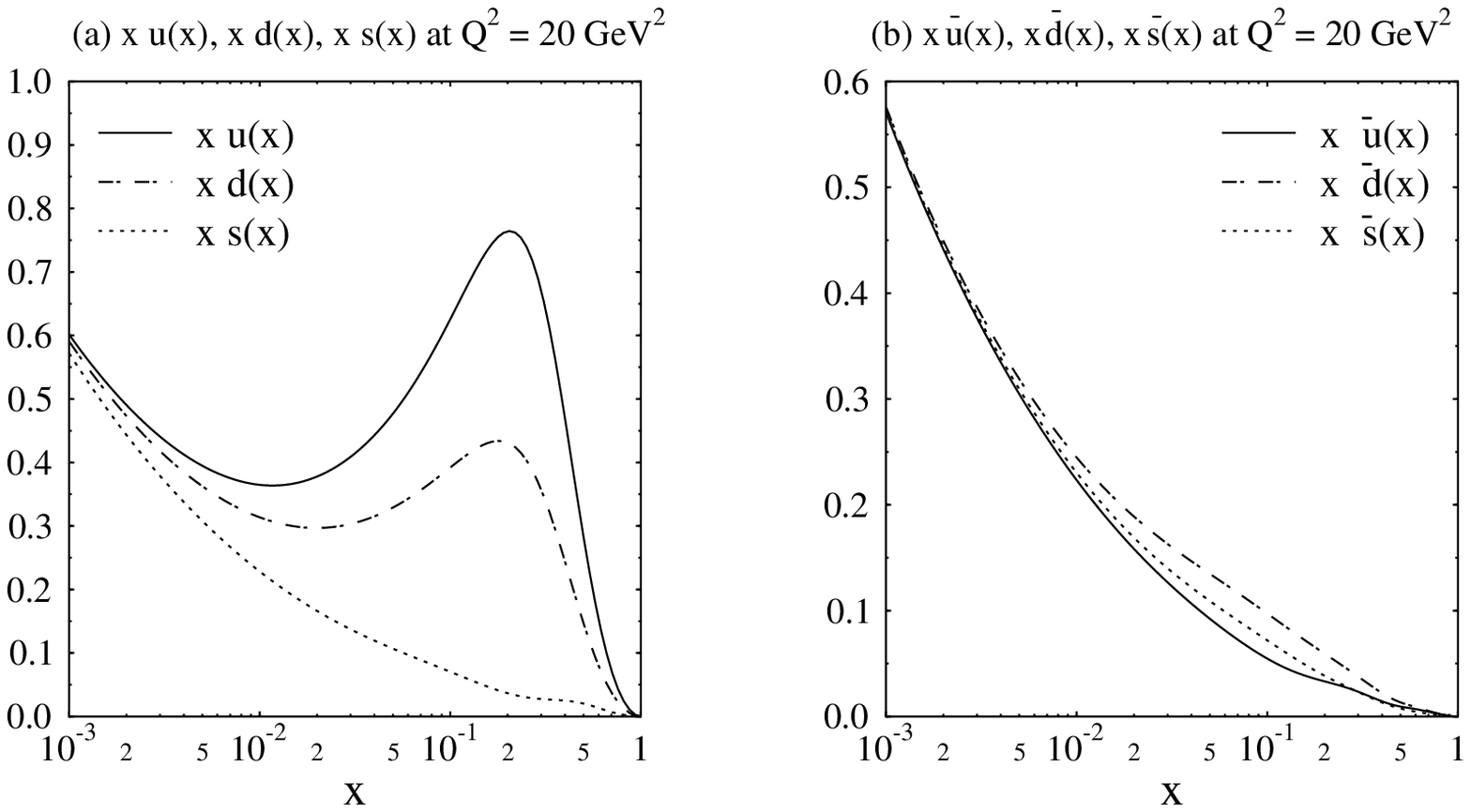}
\end{center}
\vspace*{-0.5cm}
\renewcommand{\baselinestretch}{1.20}
\caption{The theoretical predictions of the SU(3) CQSM
for the unpolarized quark and antiquark distribution functions
with each flavor at the energy scale of $Q^2 = 20 \,\mbox{GeV}^2$.
\label{unpoluds}}
\end{figure}

Next, in Fig.~\ref{na51}, the theoretical 
predictions for the ratio $\bar{d} (x) / \bar{u} (x)$ at 
$Q^2 = 30 \,\mbox{GeV}^2$ are compared with the corresponding
E866 data as well as the old NA51 data. This ratio turns out to
be a little sensitive to the flavor SU(3) generalization of the model.
It is found that the SU(3) version of the CQSM well reproduces the 
qualitative tendency of the E866 data for this ratio, although the
magnitude itself is a little overestimated.
To reveal the reason of this overestimation,
it may be interesting to compare the magnitudes of quark and
antiquark distributions themselves. Shown in Fig.~\ref{unpoluds}
are the predictions of the SU(3) CQSM for the unpolarized quark
and antiquark distribution functions with each flavor at
$Q^2 = 20 \,\mbox{GeV}^2$. Of special interest here are the
magnitudes of $\bar{u}(x), \bar{d}(x)$ and $\bar{s}(x)$.
The model predicts that
\begin{equation}
 \bar{d}(x) \ > \ \bar{s}(x) \ > \ \bar{u}(x) ,
\end{equation}
while the standard MRST \cite{MRST98} or CTEQ fit \cite{CTEQ00}
says that
\begin{equation}
 \bar{d}(x) \ > \ \bar{u}(x) \ > \ \bar{s}(x) .
\end{equation}
Undoubtedly, the magnitudes of $\bar{u}$-distribution as
compared with the other two flavors seems to be underestimated a
little too much by some reason.

As repeatedly emphasized, a quite unique feature of the SU(2) 
CQSM is that it predicts sizably large isospin asymmetry not only 
for the unpolarized sea-quark distributions but also for the 
longitudinally polarized ones. A  natural question is what the 
predictions of the flavor SU(3) CQSM is like. We have already shown 
that the predictions of these two models are not largely different 
for the isospin asymmetry of the unpolarized seas, for instance, 
for $\bar{d} (x) - \bar{u} (x)$.
To confirm it again but here
in comparison with the case of polarized distributions,
we show in Fig.~\ref{asymsea} the 
theoretical predictions for $\bar{d} (x) - \bar{u} (x)$ evaluated 
at $Q^2 = 0.88 \mbox{GeV}^2$ in comparison with Bhalerao's
semi-theoretical prediction for reference \cite{B01}.
The solid and dashed-dotted curves in 
Fig.~\ref{asymsea}(a) are the predictions of the SU(3) CQSM
obtained with and 
without the $\Delta m_s$ corrections, while the dotted curve is 
the prediction of the SU(2) CQSM. We recall that Bhalerao's prediction 
shown by the dotted curve is obtained based on what-he-call the 
statistical quark model, which is based on some statistical assumptions
on the parton distributions while introducing several experimental 
information. As one can see, all the four curves are more or less 
degenerate and they are all qualitatively consistent with the 
magnitude of isospin asymmetry of $\bar{d}$-sea and $\bar{u}$-sea 
observed by the NMC measurement. 
On the other hand, Fig.~\ref{asymsea}(b) shows the similar
analysis for the 
longitudinally polarized sea-quark distributions 
$\Delta \bar{u} (x) - \Delta \bar{d} (x)$. 
The meaning of the curves are all similar as in
Fig.~\ref{asymsea}(a). One finds 
that the magnitude of $\Delta \bar{u} (x) - \Delta \bar{d} (x)$ is 
fairly sensitive to the flavor SU(3) generalization of the CQSM, 
or more precisely, to the difference between the dynamical assumptions 
of the two models. (This provides us with one of the few exceptions 
to our earlier statement that $u, d$-flavor dominated observables 
are generally insensitive to it.) The sign of $\Delta \bar{u} (x) 
- \Delta \bar{d} (x)$ remains definitely positive but its magnitude 
is reduced by nearly a factor of 2 when going from the SU(2) model
to the SU(3) model the chiral limit 
($\Delta m_s = 0$). As was conjectured in \cite{W02}, the inclusion
of the SU(3) symmetry breaking corrections partially cancels this 
reduction and works to pull back the prediction of the SU(3) model 
toward that of the SU(2) model. Still, the final prediction of 
the SU(3) CQSM is fairly small as compared with that of the SU(2) one
although it is not extremely far from the prediction of 
Bhalerao's statistical model \cite{B01}.

\begin{figure}[htb] \centering
\begin{center}
 \includegraphics[width=15.0cm]{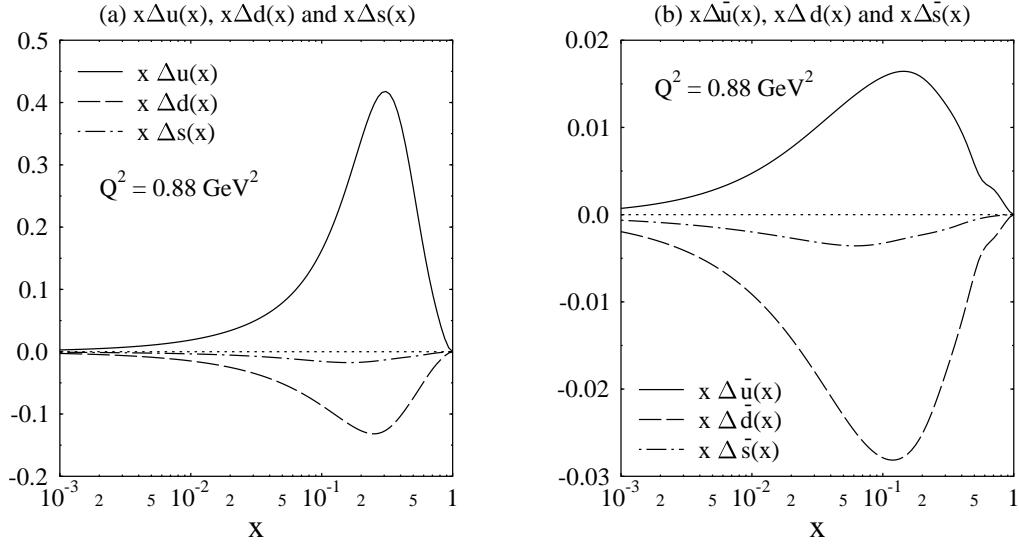}
\end{center}
\vspace*{-0.5cm}
\renewcommand{\baselinestretch}{1.20}
\caption{The theoretical predictions of the SU(3) CQSM for the
longitudinally polarized quark and antiquark distribution
functions with each flavor at the scale of
$Q^2 = 0.88 \,\mbox{GeV}^2$. \label{lgpoluds}}
\end{figure}

One may be also interested in the signs and the relative order of
the absolute magnitudes of
$\Delta \bar{u}(x), \Delta \bar{d}(x)$ and $\Delta \bar{s}(x)$
themselves.
We show in Fig.~\ref{lgpoluds} the theoretical predictions of
the SU(3) CQSM for the longitudinally polarized quark and
antiquark distributions with each flavor at the energy scale of
$Q^2 = 0.88 \,\mbox{GeV}^2$. In addition to that the
model reproduces the well-established fact $\Delta u(x) > 0$ and
$\Delta d(x) < 0$, it also predicts that $\Delta \bar{u}(x) > 0,
\Delta \bar{d}(x) < 0$ and $\Delta \bar{s}(x) < 0$ with
\begin{equation}
 \vert \Delta \bar{d}(x) \vert \ > \ 
 \vert \Delta \bar{u}(x) \vert \ > \  
 \vert \Delta \bar{s}(x) \vert .
\end{equation}
We point out that these predictions of the SU(3) CQSM are
qualitatively consistent with those of Bhalerao's statistical
quark model except for the fact that he assumes $\Delta s(x) =
\Delta \bar{s}(x)$, while the SU(3) CQSM indicates that
\begin{equation}
 \vert \Delta s(x) \vert \ \gg \ \vert \Delta \bar{s}(x) \vert .
\end{equation}

Summarizing the predictions of the two versions of the CQSM for 
the light-flavor sea-quark asymmetry, both turn out to give equally 
good explanation for the shape and magnitude of spin-independent 
distribution $\bar{d} (x) - \bar{u} (x)$. The situation is a little 
different for the longitudinally polarized sea-quark distributions.
Although the sign of $\Delta \bar{u} (x) - \Delta \bar{d} (x)$ is 
definitely positive in both models, the SU(2) CQSM predicts that
\begin{equation}
 | \bar{u} (x) - \bar{d} (x) | \ < \  
 | \Delta \bar{u} (x) - \Delta \bar{d} (x) | ,
\end{equation}
while the SU(3) model gives 
\begin{equation}
 | \bar{u} (x) - \bar{d} (x) | \ \simeq \  
 | \Delta \bar{u} (x) - \Delta \bar{d} (x) | ,
\end{equation}
or $| \Delta \bar{u} (x) - \Delta \bar{d} (x) |$ is slightly small 
than $| \bar{u} (x) - \bar{d} (x) |$. Still, the sizably large
isospin asymmetry of the spin-dependent sea-quark distributions is 
a common feature of the two versions of the CQSM.
(It is interesting to remember that large flavor and spin
asymmetry of the nucleon sea is also predicted by the instanton
model \cite{DK93},\cite{DKZ93}.) We think that this fact is worthy 
of special mention. The reason becomes clear if one compares the 
predictions of the CQSM with those of the naive meson cloud 
convolution model. As is widely known, the NMC observation 
$\bar{d} (x) - \bar{u} (x) > 0$ in the proton can be explained 
equally well by the CQSM and by the meson cloud
model \cite{K98},\cite{GP01},\cite{Ram97}. A simple 
intuitive argument, however, indicates that the latter model would 
generally predict both of $\Delta \bar{u} (x) $ and 
$\Delta \bar{d} (x)$ is small.
This is because the lightest meson, i.e. 
the pion has no spin and that the effect of heavier meson is expected 
to be less important. Actually, the situation seems a little more 
complicated. In a recent paper, Kumano and Miyama estimated the 
contribution of $\rho$-meson to the asymmetry $\Delta \bar{u} (x) 
- \Delta \bar{d} (x)$ and found that it is slightly
negative \cite{KM02}. On the other hand, Fries et al. argued that
a large positive $\Delta \bar{u} (x) - \Delta \bar{d} (x)$, as
obtained in the CQSM, can be obtained from $\pi N$-$\sigma N$
interference-type contributions in the meson cloud picture
\cite{FSW02}. Undoubtedly, for drawing
a definite conclusion within the framework of the meson 
cloud model, more exhaustive studies of possibly important
Feynman diagrams are necessary.
This should be contrasted with the prediction of the 
CQSM.
Since there is little arbitrariness in its theoretical framework, 
its prediction once given is one and only in nature and cannot be 
easily modified.
Both of the CQSM and the meson cloud convolution model give 
equally nice explanation for the novel isospin asymmetry for the 
unpolarized sea-quark distributions, so that one might have naively
thought that they are two similar models containing basically
the same physics.
In fact, a commonly important ingredients of the two models are 
the Nambu-Goldstone pions resulting from the spontaneous chiral 
symmetry breaking of QCD vacuum. However, a lesson learned from
the above consideration of the isospin asymmetry for the
spin-dependent sea-quark distributions is that it is not necessarily
true. An interesting question is what 
makes a marked difference between these two models. In our opinion, 
it is a strong correlation between spin and isospin quantum numbers 
embedded in the basic dynamical assumption of the CQSM, i.e. 
the hedgehog ansatz. We recall that we have long known one example
in which the difference of these two models makes more profound
effect \cite{W89},\cite{W90},\cite{WY91},\cite{WW00B}.
It is just the problem 
of quark spin fraction of the nucleon. Is there any simple 
and convincing explanation of this nucleon spin puzzle within the 
framework of the meson cloud model? The answer is no, to our knowledge.
On the other hand, assuming that the 
dynamical assumption of the CQSM is justified in nature, it gives 
quite a natural answer to the question why the quark spin fraction 
of the nucleon is so small. In fact, according to this model,
a nucleon is a bound state of quarks and 
antiquarks moving in the rotating mean-field of hedgehog shape.
Because of the collective rotational motion, it happens that a 
sizable amount ($\sim 65 \%$) of the total nucleon spin is carried 
by the orbital angular momentum of quark and antiquark fields.
We conjecture that the cause of the {\it simultaneous large violation}
of the {\it isospin asymmetry} for both the {\it spin-independent}
and {\it spin-dependent sea-quark distributions} can also be traced
back to the strong {\it spin-isospin correlation} generated by the
formation of the hedgehog mean field.

\section{Summary and Conclusion}

In summary, several theoretical predictions are given for the
light-flavor quark and antiquark distribution functions in the
nucleon on the basis of the flavor SU(3) CQSM.
Its basic lagrangian is a straightforward generalization of the
corresponding SU(2) model except for the presence of sizably
large SU(3) symmetry breaking term, which comes from the appreciable
mass difference $\Delta m_s$ between the $s$-quark and
the $u,d$-quarks. As explained in the text, this SU(3) symmetry
breaking effect is treated by using a perturbation theory in the
mass parameter $\Delta m_s$.
We have shown that the SU(3) CQSM can give several unique predictions 
for the strange and antistrange quark distributions in the nucleon 
while maintaining the success previously obtained in the flavor SU(2) 
version of the CQSM for $u,d$-flavor dominated observables.
For instance, it predicts a sizable amount of particle-antiparticle
asymmetry for the strange-quark distributions.
Its predictions for the distributions $s (x) - \bar{s} (x)$ and 
$s (x) / \bar{s} (x)$ at $Q^2 = 20 \,\mbox{GeV}^2$ are shown to be
consistent with the corresponding phenomenological information
given by Barone et al. and by CCFR group at least qualitatively.
As expected, the magnitudes 
of $s (x)$ and $\bar{s} (x)$ turn out to be very sensitive 
to the SU(3) symmetry breaking effects. We showed that the theoretical 
predictions for $x \bar{s} (x)$ and $x \bar{s} (x)$ at
$Q^2 = 4 \,\mbox{GeV}^2$ and $Q^2 = 20 \,\mbox{GeV}^2$ are
qualitatively consistent with the CCFR data 
after taking account of the SU(3) symmetry breaking effects.
The particle-antiparticle asymmetry of the strange quark distributions 
are even more profound for the spin-dependent distributions than for 
the unpolarized distributions. Our theoretical analysis strongly
indicates that the negative (spin) polarization of the strange 
quarks, i.e. the fact that $\Delta s (x) + \Delta \bar{s} (x) < 0$,
as suggested by the LSS fit as well as many other phenomenological
analyses, comes almost solely from the $s$-quark and the polarization
of $\bar{s}$-quark is very small.
The model gives interesting predictions also for 
the isospin asymmetry of the $\bar{u}$- and $\bar{d}$-quark
distributions. We had already 
known that the flavor SU(2) CQSM gives a natural explanation of 
the NMC observation, i.e. the excess of $\bar{d}$-sea over the 
$\bar{u}$-sea in the proton. In the present investigation, we have
confirmed that this favorable aspect of the SU(2) CQSM is just
taken over by the SU(3) CQSM and that they in fact give nearly the 
same predictions for the magnitude of the asymmetry 
$\bar{d} (x) - \bar{u} (x)$. On the other hand, we find that the 
predictions of the two models for the isospin asymmetry of the 
longitudinally polarized sea-quark distributions are a little 
different. Both models predicts that $\Delta \bar{u} (x) 
- \Delta \bar{d} (x) > 0$, but the magnitude of asymmetry is reduced 
by a factor of about 0.6 when going from the SU(2) model to the 
SU(3) one.
Still, a sizably large isospin asymmetry of the spin-dependent 
sea-quark distributions is a common prediction of both versions of 
the CQSM, and it should be compared with the unsettled 
situation in the meson-cloud convolution models.
In our opinion, the physical origin of the simultaneous violation 
of the isospin symmetry for the spin-independent and spin-dependent 
sea-quark distributions may be traced back to the strong correlation 
between spin and isospin embedded in the hedgehog symmetry of soliton 
solution expected to be realized in the large $N_c$ limit of QCD.
What should be emphasized here is 
another consequence of the hedgehog symmetry embedded in the CQSM.
It has long been recognized that, according to this model, only about 
$35 \,\%$ of the total nucleon spin is due to the intrinsic quark
spin and the remaining $65 \,\%$ is borne by the orbital angular
momentum of quark and antiquark fields.
We emphasize that this is a natural consequence 
of the nucleon picture of this model, i.e. ``rotating hedgehog''.
Unfortunately, unresolved role of gluon fields, especially the 
role of $U_A (1)$ anomaly makes it difficult to draw a definite 
conclusion on this interesting but mysterious problem.
In this respect, more 
thorough study of simpler problem, i.e. the possible isospin asymmetry 
of longitudinally polarized sea-quark distributions may be of
some help to test the validity of the basic idea of the 
soliton picture of the nucleon. At any rate, an important lesson
learned from our whole analyses is that the
{\it spin and flavor dependencies} of
{\it antiquark distributions} in the nucleon are very sensitive to 
the nonperturbative dynamics of QCD. To reveal this interesting 
aspect of baryon structures, it is absolutely necessary to carry out 
{\it flavor} and {\it valence plus sea quark decompositions} of
the parton distribution functions.
We hope that this expectation will soon be 
fulfilled by various types of semi-inclusive DIS scatterings as well 
as neutrino-induced reactions planned in the near future.

\begin{acknowledgments}
This work is supported in part by a Grant-in-Aid for Scientific
Research for Ministry of Education, Culture, Sports, Science
and Technology, Japan (No.~C-12640267)
\end{acknowledgments}

\bibliographystyle{unsrt}
\bibliography{ref1}

\end{document}